\documentclass[usenatbib]{mn2e}
\usepackage{appendix}
\usepackage{graphicx}
\usepackage{float}
\usepackage{cite}
\usepackage{amssymb,amsmath}
\usepackage[mathscr]{eucal}
\usepackage[usenames]{color}

\input{macro1.mac}

\begin{document}
\newcommand{\risa}[1]{\textcolor{red}{\bf #1}}

\title[Formation of disk galaxies in preheated media] 
{Formation of disk galaxies in preheated media: a preventative feedback model} 
\author[] {Yu Lu$^{1}$\thanks{E-mail: luyu@stanford.edu}, 
H.J. Mo$^{2}$, 
Risa H. Wechsler$^{1}$
  \\
  $^1$ Kavli Institute for Particle Astrophysics and Cosmology,
  Physics Department, \\ and SLAC National Accelerator Laboratory,
  Stanford University, Stanford, CA 94305, USA
  \\
  $^2$ Department of Astronomy, University of Massachusetts, Amherst
  MA 01003-9305, USA}


\date{}

\maketitle

\begin{abstract}
We introduce a semi-analytic galaxy formation model implementing a
self-consistent treatment for the hot halo gas configuration and the assembly of 
central disks. 
Using the model, we explore a preventative feedback model, in which 
the circum-halo medium is assumed to be preheated up 
to a certain entropy level by early starbursts or other processes, 
and compare it with an ejective feedback model, in which baryons are first accreted into dark matter halos 
and subsequently ejected out by feedback. 
The model demonstrates that when the medium is preheated to an entropy 
comparable to the halo virial entropy the baryon accretion can be largely reduced 
and delayed. In addition, the preheated medium can establish an extended 
low density gaseous halo when it accretes into the dark matter halos, and result in  
a specific angular momentum of the cooling gas large enough to form central disks 
as extended as those observed. 
Combined with simulated halo assembly histories,  
the preventative feedback model can reproduce remarkably well 
a number of observational scaling relations. These include the cold baryon 
(stellar plus cold gas) mass fraction-halo mass relations, star formation histories, 
disk size-stellar mass relation and its evolution, and the 
number density of low-mass galaxies as a function of redshift.
In contrast, the conventional ejective feedback model fails to reproduce these observational trends.   
Using the model, we demonstrate that 
the properties of disk galaxies are closely tied to the thermal state of 
hot halo gas and even possibly the circum-halo medium, 
which suggests that observational data for the disk properties and circum-galactic 
hot/warm medium may jointly provide interesting constraints for galaxy formation models.  
\end{abstract}

\section{Introduction}\label{sec:introduction}

Galaxy formation remains one of the most challenging problems in astrophysics
largely because galaxy formation involves many complicated baryonic processes, 
which are still poorly understood \citep[e.g.][]{Fall2002, Mo2010, Silk2012}. 
The early theories, pioneered by \citet{Binney1977, Rees1977, Silk1977}, and \citet{White1978}, 
used a cooling rate argument to place an upper limit on the masses of galaxies. 
Soon after that, people realized that some form of feedback has to be involved 
in galaxy formation to reduce the star formation efficiency, especially in low-mass systems \citep[e.g.][]{Dekel1986}. 
In the conventional picture of galaxy formation, 
all baryons are first accreted into dark matter halos and a substantial fraction  is subsequently heated  
and ejected out of the halo by feedback processes. 
Various sophisticated models based on this general assumption have been built and 
explored extensively \citep{Croton2006, Bower2006, Somerville2008, Lu2011a}. 
However, no galaxy formation model has been able to reproduce 
the full range of the most important observational constraints with satisfaction. 
To date, models either poorly reproduce observational data or are based on physically 
implausible assumptions \citep[see e.g.][]{Mutch2013, Lu2013a}.  
It is important, at this point, to explore models beyond the standard assumption of full baryon accretion 
and ejective feedback. 
In this paper, we propose a new model for a preventative feedback 
process based on an assumption that the circum-galactic medium is preheated to 
a certain level of entropy by early feedback processes. 
We use a new semi-analytic model (SAM) to explore the basic consequences of the assumption 
of preheating on galaxy formation and 
contrast the model with the conventional assumption 
of full baryon accretion and ejective feedback in low-mass halos. 

A great wealth of observational data have become available to test
models of galaxy formation.  With multi-wavelength surveys, such as
the SDSS \citep[][]{York2000} and GAMA \citep[][]{Driver2011} surveys,
the luminosity/stellar mass functions of the local galaxy population
can now be measured to high accuracy \citep[e.g.][]{Blanton2005,
  Moustakas2013}.  Surveys in radio bands, such as HIPASS
\citep{Meyer2004, Zwaan2004} and ALFALFA \citep{Giovanelli2005,
  Giovanelli2005a}, have also made it possible to estimate the cold
gas mass function of local galaxies, providing a complete census of
the cold baryonic contents of present-day galaxies.  Combined with the
halo mass function predicted by the current $\Lambda$CDM model, these
data can be used to statistically establish connections between
galaxies and their host dark matter halos \citep{Yang2003, Conroy2006,
  Yang2012, Yang2013, Behroozi2010, Behroozi2012, Moster2013, Lu2013,
  Reddick2013}. This type of study has produced important results for
the luminosity/stellar mass - halo mass relation, the cold gas mass -
halo mass relation, and the star formation and stellar mass assembly
histories in halos of different masses.  An important characteristic
trend is that the fraction of cold baryons (stars and cold gas)
decreases rapidly with decreasing halo mass for halos with virial
masses $M_{\rm vir}<10^{12}\msun$ \citep{Yang2003, Yang2008,
  vandenBosch2003, Behroozi2010, Behroozi2012, Papastergis2012}, which
provides stringent constraints on the star formation and assembly
processes modeled in SAMs \citep{Croton2006, Bower2006,
  Somerville2008,Lu2012, Lu2013a} and in hydrodynamical simulations
\citep[e.g.][]{Dave2013, Hopkins2013}.

Another important observational constraint, whose constraining power has not yet 
been widely exploited, comes from the sizes of galaxies. 
With the advent of deep surveys from both ground-based 
and space telescopes,  the galaxy luminosity/stellar mass 
functions have been estimated to $z\sim8$ \citep[e.g.][]{Bradley2012, Finkelstein2012, Oesch2012, Yan2012, McLure2013}, 
allowing us to study the star formation and assembly histories
of galaxies over the cosmic time. In addition, high quality imaging data
from the HST also permits studies of the evolution of  galaxy morphology 
out to $z\sim2$ \citep[e.g.][]{Trujillo2006, Cassata2013, Patel2013a, Patel2013, vanderWel2014}.

Based on the conventional assumption that all baryons are accreted into dark matter halos, 
models predict that the cooling efficiency in low-mass halos is always much higher than the star
formation efficiency in observed galaxies \citep{Thoul1996,
Silk1997, Kennicutt1998}. 
Assuming strong feedback, a 
number of models have successfully reproduced the galaxy mass 
function or luminosity function observed in the local universe \citep{Croton2006,
Bower2006, Somerville2008, Lu2013b}.  
However, simultaneously reproducing the observed
number density of galaxies at different redshifts, the baryon mass--halo
mass relation, and the size--mass relation remains an extremely challenging 
problem in galaxy formation \citep[see e.g. discussion in][]{Weinmann2012, Lu2013a}. 
The formation histories of galaxies, especially those with masses 
equal to or lower than that of the Milky Way,
are also hard to reproduce. Observed low-mass field galaxies 
are typically star forming, maintaining a specific star rate  
${\rm sSFR}\approx 0.5 {\rm Gyr}^{-1}$ all the way to the 
present day \citep{Moustakas2013}, while star formation 
feedback in theoretical models is typically more effective in 
lower mass halos, resulting in the predicted star formation rates 
in faint galaxies that are too low  \citep{Weinmann2012, Wang2012, Lu2013a}. 
In addition, low-mass halos are predicted to form stars very efficiently
at high redshift, producing too many  low-mass  
galaxies at $z>2$ to match observations \citep{Weinmann2012, Lu2013a}.

The extended size of observed disk galaxies has been another long-standing challenge for modeling galaxy formation. 
Early hydrodynamical simulations 
showed that the cooling gas forming the central galaxy in a halo
has too low angular momentum to produce an extended disk \citep[e.g.][]{Katz1991, Navarro1991, Steinmetz1999}.
This angular momentum``catastrophe'' has been attributed partially to numerical limitations, 
and partially to uncertainties in modeling baryonic processes such as feedback \citet{Fall2002, Maller2002}. 
Analytic and semi-analytic models normally assume that the baryons are initially mixed 
with the dark matter and share the same specific angular momentum, $j$. 
Even when the baryons cool and decouple from the dark matter to collapse on a disk, 
the material assemblies the disk still retains the same $j$ \citep[e.g.][]{Fall1980, Dalcanton1997, Mo1998, Somerville2008a}. 
Assuming the baryonic $j$ is conserved, such models can reproduce 
both the zero point and the slope of the observed spiral-galaxy 
$j_*-M_*$ relation \citep{Dutton2012, Romanowsky2012}.
However, this conventional assumption is implausible since the collapse of 
dark matter and cooling of gas are governed by different physical processes 
and occur on different scales in space and time. 
Cooling preferentially happens in the inner regions of the halo, 
while the outer regions remain gaseous and has less cooling.
As demonstrated by \citet{Dutton2012}, in conventional models, 
the concentrated hot gas distribution leads to rapid gas cooling in the halo center, 
where the specific angular momentum is low, and, hence, results in too low 
angular momentum of the disk. 
Therefore, the ``angular momentum catastrophe" also exists in 
semi-analytic models if cooling and angular momentum distribution of halo gas 
are treated self-consistently.

All of these issues could be centered on our understanding of feedback processes, 
which is one of the biggest uncertainties in modeling galaxy formation.
Most models investigated so far  assume that feedback is ejective, in the sense that supernova 
explosions associated with stellar evolution ejects cold gas from 
the disk, thereby reducing subsequent star formation. However, it is unclear how effective 
such feedback is in reality. Detailed simulations capable of 
resolving the interface of the SN-driven super-bubbles and the
ISM have shown that the feedback is inefficient in driving gas 
out of a galaxy because of the rapid development of 
Rayleigh-Taylor instabilities \citep{MacLow1999, Krumholz2012}. 
Yet, a high fraction of supernova energy is required to be 
coupled with the ISM in order to explain the faint-end of 
galaxy luminosity function, the low-mass end of 
the HI mass function \citep{Lu2013a}, and the evolution of 
the stellar mass function \citep{Mutch2013}.
Moreover, \citet{Henriques2013} found that, even if supernova
is effective in ejecting gas from galaxies, the ejected gas 
is required to follow a particular schedule to reincorporate 
into the halo later in order to  simultaneously match the observed 
galaxy luminosity functions at multiple redshifts. Recent 
hydrodynamical simulations suggest that additional energy 
from radiation pressure associated with massive stars 
may be able to provide sufficient amounts of energy to 
reproduce the observed low baryon mass fractions in 
low-mass halos \citep{Stinson2013,Dave2013,Hopkins2013}. 
However, how the feedback governs the budget of 
the baryonic matter in these simulations is still unclear. 
Observationally, star forming galaxies at $z\lesssim2$ only 
show evidence of outflows with a mass-loading factor 
$\sim2$ \citep{Bouche2012, Newman2012}, which does not 
seem to be sufficient to explain the low baryon mass fraction 
in low-mass halos \citep{Papastergis2012, Lu2013a}. 

In this paper, we propose a new galaxy formation model 
that implements a preventative scenario of feedback. 
We assume that the intergalactic gas is heated to some 
finite entropy before it is accreted into dark matter haloes, 
motivated by processes that have been suggested in the literature, 
such as preheating by supernova/AGN winds \citep{Mo2002,Mo2004},  
by gravitational pancaking \citep{Mo2005, Lu2007},  by 
blazar heating \citep{Pfrommer2012}, and by intergalactic 
turbulence \citep{Zhu2011}. The enhanced entropy affects 
galaxy formation in two ways. First, the baryon fraction that can 
collapse into low mass halos is strongly reduced 
\citep{Mo2002, vandenBosch2003, Oh2003, McCarthy2004, Scannapieco2004, Lu2007}. 
Second, when the entropy of the pre-collapsed gas is higher than
would be generated by accretion shocks, the halo gas is expected to 
develop an extended density distribution 
\citep{Mo1996, Mo2002, Maller2004, Kaufmann2009, Fang2013},
thereby affecting where and when halo gas cools to fuel 
the central disk. Using a semi-analytic model built upon 
simulated halo accretion histories, we examine in detail 
the disk size evolution, mass assembly and star formation 
histories of central galaxies hosted  by halos with masses 
$M_{\rm vir}<10^{12}\msun$ at the present time in the
preheating scenario. We also compare the results with those predicted 
by an ejective feedback model and with observational data.


The paper is organized as follows. In \S \ref{sec:model}, we
introduce the physics of how a preheated circum-halo medium affects galaxy formation 
and describe the implementation of the model. We show the predictions of a typical 
ejective feedback model and our preventative feedback model, and compare the model predictions 
with observations in \S \ref{sec:results}. 
In \S \ref{sec:conclusion}, we summarize our results and discuss their implications. 
We also describe the detailed implementations of model recipes for  
reionization, star formation in galaxy disks, and stellar mass loss due to stellar evolution 
in the Appendixes. 
Throughout the paper, we use a
$\Lambda$CDM cosmology with $\Omega_{\rm M,0} = 0.27$,
$\Omega_{\Lambda,0} = 0.73$, $\Omega_{\rm B,0} = 0.044$, $h = 0.70$,
$n = 0.95$, and $\sigma_8 = 0.82$. 

\section{The Model}\label{sec:model}

To study the impact of preheated circum-halo gas to the formation of
disk galaxies, we develop a new semi-analytic model, which follows
realistic halo mass assembly histories and includes the most important
physical processes that are generally implemented in galaxy formation
models.  Our model consists of the following parts: 1) halo mass
accretion histories and density profiles; 2) gas accretion and
distribution in dark matter halos; 3) radiative cooling of halo gas
and formation of galaxy disks; 4) star formation and supernova (SN)
feedback in galaxy disks.  The model follows these processes and makes
predictions for the evolution of central galaxies from an early time
to the present day.  More importantly, the prescriptions include the
key physics that allows us to model galaxy formation based on
different assumptions for the entropy level of the circum-halo gas.
The reason for us to develop this completely new model, instead of
using our existing model \citep{Lu2013a, Lu2013b} is because the
present study focuses on experimenting with new physics rather than
exploring the parameter space, which is the main task for our
previously published models.  In the following, we describe the model
and highlight the model prescriptions that are newly implemented for
this paper.

\subsection{Formation history and structure of dark matter halos}
\label{sec:model_halo}  

\setcounter{figure}{0}
\begin{figure*}
\centering
\includegraphics[width=0.45\textwidth]{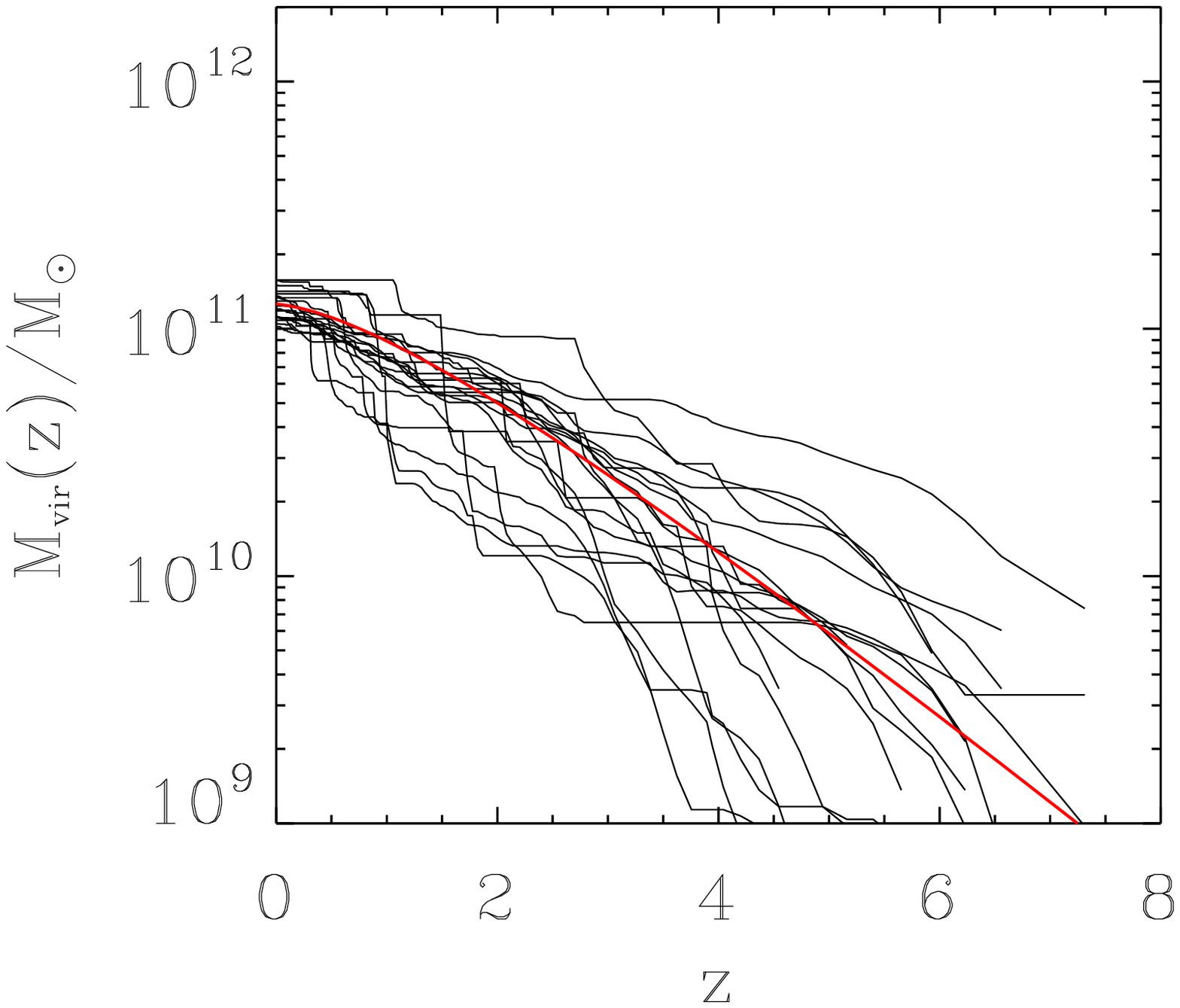}
\includegraphics[width=0.45\textwidth]{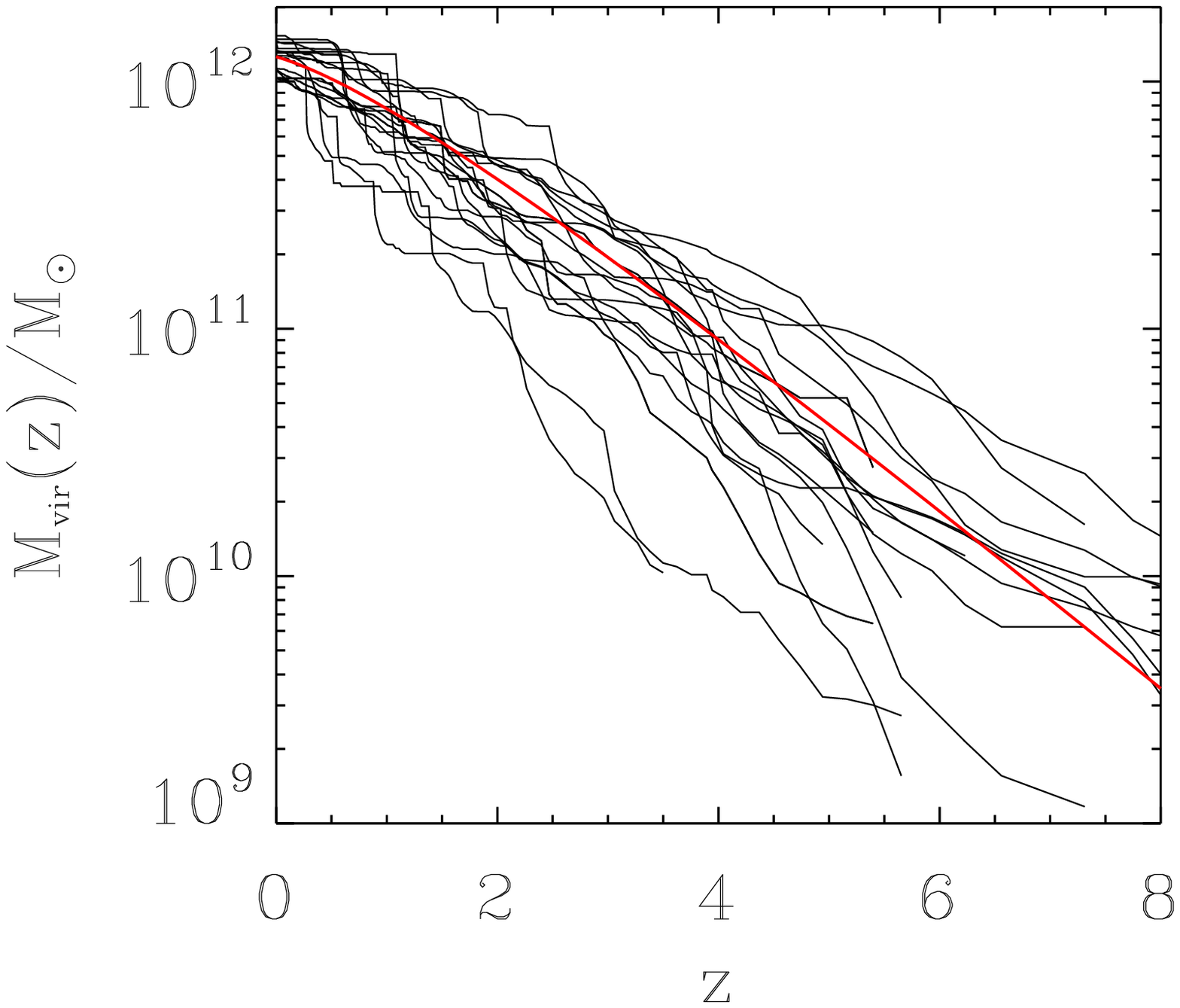}
\caption{Mass assembly histories (MAHs) of dark matter 
halos with a final ($z=0$) virial mass in the range of $10^{11}-10^{11.2}\msun$ (left) 
and in the range of $10^{12}-10^{12.2}\msun$ (right). The black lines show 20 MAHs 
randomly selected from the Bolshoi simulation. The red lines are a 
smoothed version of the MAHs using the McBride model 
(Eq. \ref{equ:mah}) with $\alpha=0.8$, $\eta=0.9$ 
for $M_{\rm vir,0}=10^{11.1}\msun$ halos, and
$\alpha=0.6$, $\eta=0.9$ for $M_{\rm vir,0}=10^{12.1}\msun$ halos.
}
\label{fig:mah}
\end{figure*}

Our model follows the main-branch mass assembly history (MAH) 
of dark matter halos. Any secondary progenitors and satellite 
galaxies associated with subhalos are ignored. 
For halos hosting a galaxy with a stellar mass comparable to or lower
than that of the Milky Way, 
this is a good approximation because the stellar mass in 
these halos is assembled mainly through {\it in situ} star formation 
in the main-branch progenitor rather than through mergers 
\citep{Yang2012, Yang2013, Behroozi2012, Lu2013}.   
We adopt realistic MAHs extracted directly from 
a $N$-body cosmological simulation, the Bolshoi simulation 
\citep{Klypin2011}.
Dark matter halos are identified at 180 time steps 
from $z=14$ to $z=0$ using the Rockstar halo finder 
\citep{Behroozi2013a}, and halo merger trees are constructed by 
linking a halo with all its progenitors using the Consistent 
Tree method \citep{Behroozi2013}. 
The halo mass assembly histories provide enough mass and time resolution 
to study the formation of central disk galaxies in the halo mass range of $10^{10}-10^{12}\msun$. 
For this paper, 
we randomly select a large number of halos with masses 
and redshifts that are relevant to our study. 
Figure \ref{fig:mah} shows two  
random subsets of the MAHs for halos with a present-day 
virial mass in the range of $10^{11.0}-10^{11.2}\msun$ or $10^{12.0}-10^{12.2}\msun$. 
As one can see, the simulated MAHs for 
a given final halo mass have quite large variations.  
For comparison, we also show a smoothed, average accretion history
based on the fitting formula proposed by \citet{McBride2009},
\begin{equation}\label{equ:mah}
M_{\rm vir}(z) = M_{\rm vir,0} \left( 1+z\right)^{\alpha} \exp(-\eta z),
\end{equation}
where the normalization $M_{\rm vir, 0}$ is the halo 
mass at $z=0$ and $\alpha$ and $\eta$ are 
parameters determining the shape of the MAH. 
The values of the two parameters adopted in the plot 
are $\alpha=0.6$ and $\eta=0.9$ for $M_{\rm vir, 0}\approx10^{12.1}\msun$ 
halos, and $\alpha=0.8$ and $\eta=0.9$ for $M_{\rm vir,0}\approx10^{11.1}\msun$ halos.
 
At any redshift, the density distribution within a dark matter 
halo is assumed to be spherically symmetric and to follow 
a NFW profile \citep{Navarro1996}:
\begin{equation}\label{equ:NFWprofile}
\rho(r) = {4\rho_{\rm s} \over (r/r_{\rm s}) (1+r/r_{\rm s})^2}\,,
\end{equation}
where $\rho_{\rm s}$ and $r_{\rm s}$ are the characteristic density 
and radius of the halo. 
The halo concentration is defined as the ratio of the virial 
radius to the characteristic radius as $c=\rvir/r_{\rm s}$. 
Cosmological simulations show that the halo concentration 
depends on both halo mass and redshift at which 
the halo is identified \citep[e.g.][]{Bullock2001, Eke2001, 
Zhao2003, Zhao2009, Maccio2007}. 
Here we adopt the recent simulation result 
of \citet{Prada2012} to compute $c$ for a halo 
with a given mass at a given redshift. 
In this model, the concentration of a halo increases with time. 
For a present-day $M_{\rm vir}=10^{12}\msun$ halo, 
the typical concentration is $c\approx10$,  and $c\approx 7$ at $z=2$.

\subsection{Baryon accretion of dark matter halos}
\label{sec:model_hot}

As a dark matter halo grows, a certain amount of baryonic mass 
is expected to follow dark matter to collapse into the halo. 
One of the most basic differences between our preventative model and 
normal ejective models is that in our preventative model, only a reduced fraction of baryonic matter 
can collapse into a halo, while a nearly cosmic baryon fraction of baryonic matter 
can collapse in the ejective feedback model.   
In general, we write the 
baryon accretion rate in terms of the halo mass accretion rate as
\begin{equation}
\dot{M}_{\rm acc}=f_{\rm b}\dot{M}_{\rm vir}\,,
\end{equation} 
where the coefficient $f_{\rm b}=f_{\rm acc}f_{\rm b,0}$, 
with $f_{\rm b,0}\approx 0.17$ the cosmic baryon mass fraction 
and $f_{\rm acc}$ a parameter to be determined by relevant 
baryonic processes.
Following other SAMs, we include the effect of reionization on the 
baryon accretion fraction by adopting a model proposed by \citet{Gnedin2000} and \citet{Kravtsov2004} 
(see Appendix \ref{sec:model_reionization} for details). 
For halos relevant to our study, this effect is small. 

In our preventative model, we explore the impact on galaxy formation if  
the circum-halo media is preheated to an entropy level higher than that due to 
the photoionization heating alone. In this model, we assume that the baryons 
are preheated to have an entropy $S$.  
\citet{Lu2007} simulated the accretion process of the preheated gas 
onto growing dark matter halos and found that the baryon mass fraction in a halo  
scales with the preheating entropy $S$ and 
the virial entropy of halo, $S_{\rm vir}$, as
\begin{equation}\label{equ:fb_preh}
f_{\rm b}={1 \over \left[1+\left({S/S_{\rm vir} 
\over 0.8}\right)^3\right]^{1/2}}\,,
\end{equation}
where the virial entropy is defined as 
\begin{equation}\label{equ:svir}
S_{\rm vir}={T_{\rm vir} \over n_{\rm vir}^{2/3}}\,,
\end{equation}
with $T_{\rm vir}$ the virial temperature of the halo, 
and $n_{\rm vir}$ the mean gas particle number density of  
a virialized halo assuming the cosmic baryon fraction $f_{\rm b,0}$. 
Thus, if the preheating entropy is much lower than the 
virial entropy of the halo, the baryonic matter follows 
the dark matter to collapse into the halo, resulting in 
$f_{\rm b}\approx f_{\rm b,0}$. 
If, on the other hand, the preheating entropy 
is much higher than the halo virial entropy, then  
the baryon mass fraction $f_{\rm b} \propto 
(S_{\rm vir}/S)^{3/2}$. For a constant $S$, this gives
$f_{\rm b} \propto S_{\rm vir}^{3/2} \propto M_{\rm vir}$ 
at a given redshift. 
 
The model predicts the initial fraction of baryons 
that collapse into a halo using equation (\ref{equ:fb_preh}).
We trace the evolution of the hot halo gas using 
\begin{equation}
M_{\rm hot}=f_{\rm b} M_{\rm vir} - (M_*+M_{\rm cold}+M_{\rm eject})\,,
\end{equation}
where $M_*$, $M_{\rm cold}$ and $M_{\rm eject}$ 
are the baryon masses in stars, cold gas and ejected material 
that are associated with the halo, respectively.

\subsection{Configuration of the hot halo gas}
\label{sec:model_prof}

\begin{figure*}
\begin{center}
\includegraphics[width=0.8\textwidth]{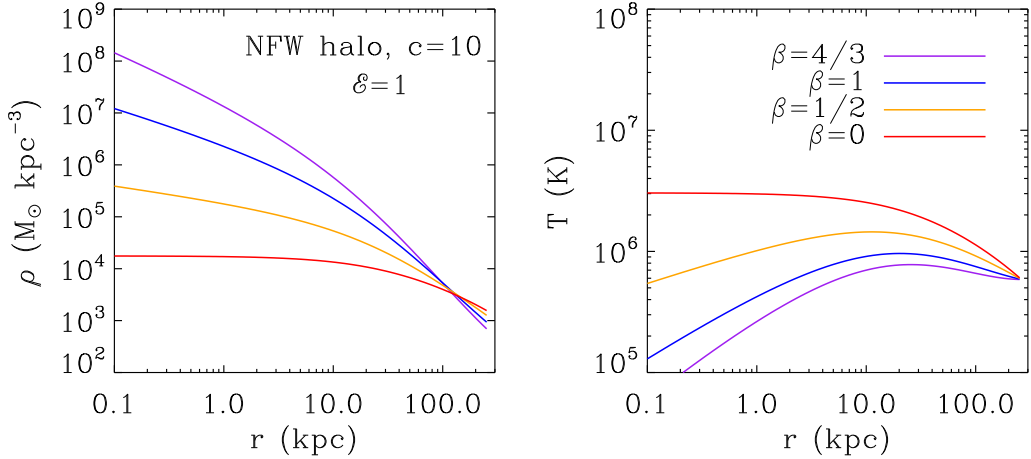}
\caption{
Density profiles (left) and temperature profiles (right) of 
the hot halo gas in hydrostatic equilibrium in a $10^{12}\msun$ 
NFW halo with $c=10$. The lines with different colors 
denote models with different logarithmic slopes for the 
entropy radial profile.  
In all the models, the total gas mass is fixed to be 
$0.17M_{\rm vir}$ and the gas entropy at the virial radius 
is normalized so that the gas temperature at the virial 
radius equals to the virial temperature of the halo
(i.e. $\mathcal{E}=1$). Note that the preheating model we explore 
in this paper corresponds to the $\beta=0$ case. 
}
\label{fig:hot_equ}
\end{center}
\end{figure*}

We model the structure of the gaseous halo under the assumption 
that the hot gas accreted into a halo is in hydrostatic 
equilibrium with the gravitational potential of the host halo.
The hot gas distribution is then solved with an assumed 
equation of state for an ideal gas: 
\begin{equation}\label{equ:eos}
p=A \rho^{\gamma},
\end{equation}
where $A$ is called the {\it adiabat}, which is a constant 
if the gas experiences an adiabatic process, and 
$\gamma={5/3}$ for a mono-atomic gas. 
The adiabat of an ideal gas can be written as 
\begin{equation} 
A={ kT \over \mu m_{\rm p} \rho^{\gamma-1}}\,,
\end{equation}
where $k$ is the Boltzmann constant, $\mu$ is the mean 
molecular weight, $m_{\rm p}$ is the mass of proton, and $T$ 
is the temperature of the gas. The adiabat
is related to the specific entropy of the gas,
\begin{equation}
A= {k \over (\mu m_{\rm p})^{\gamma}}S\,,
\end{equation}
where the specific entropy, $S$, is defined as 
\begin{equation}
S={T \over n^{2/3}}\,,
\end{equation}
with $n$ being the number density of gas particles:
$n= \rho/(\mu m_{\rm p})$. 

Analytical models and hydrodynamic simulations of halo formation
neglecting radiative cooling have shown that the radial profile of the specific 
entropy of the hot gas generated by accretion 
shocks roughly follows a power-law function of radius with a power index 
$\beta \sim 1.1$ \citep[e.g.][]{Tozzi2001, Voit2005, Lu2007}. 
When radiative cooling is included, the entropy profile can be modified. 
Using a simplified model, \citet{Bryan2000} argued that because rapid cooling always happens 
in low-entropy gas, and the high-entropy gas will expand 
adiabatically to occupy the volume of the halo, resulting in a flatter entropy profile. 
However, \citet{Tang2009} found that cooling in Milky Way 
sized halos is a runaway process in their simulations,
and the resulting entropy profile remains a power-law. 
When feedback is taken into account, the entropy profile 
can be modified further \citep{Voit2005a, McCarthy2010}.
Hydrodynamical simulations show that, when feedback is 
included, the entropy profile of the halo gas becomes shallower 
than that predicted in adiabatic simulations \citep[e.g.][]{Crain2010}. 
While observing the hot gas in low-mass halos is difficult, 
the entropy of hot halos of giant ellipticals shows a similar 
power-law profile with flattening in the center due to 
heating of feedback from central galaxies \citep{Werner2012}. 
In this paper,  we adopt a 
generic model where the entropy profile of the halo gas is 
assumed to be a power-law, 
\begin{equation}
S(r)=S_{0} \left({r\over r_{\rm vir}}\right)^{\beta}\,,
\end{equation}
or
\begin{equation}\label{equ:en_prof}
A(r)=A_{0} \left({r\over r_{\rm vir}}\right)^{\beta}\,,
\end{equation}
where $S_{0}$ and $A_{0}$ are the gas entropy and adiabat at 
the halo virial radius $r_{\rm vir}$, respectively, 
and $\beta$ determines the slope of the profile. 
Thus, setting up $\beta=0$ gives rise to an isentropic gas profile \citep[e.g.][]{Mo1996, Mo2002, Maller2004}.

To obtain the gas density profile, we solve the following 
hydrostatic equilibrium equation,
\begin{equation}\label{equ:equi}
{1 \over \rho}{{\rm d} p \over {\rm d} r} = - G{M(<r) \over r^2}\,,
\end{equation}
where $M(<r)$ is the gravitational mass within a radius $r$. 
Assuming the gravitational mass to be dominated 
by the dark matter, we obtain $M(<r)$ from the halo mass profile. 
Substituting Eqs.\,(\ref{equ:eos}) and (\ref{equ:en_prof}) into 
Eq.\,(\ref{equ:equi}), we have
\begin{equation}
{{\rm d}\rho^{\gamma-1} \over {\rm d} r} + 
{\gamma-1 \over \gamma} \left[ \beta \rho^{\gamma-1} {1\over r}+ 
G{M(<r) \over r^2} {1\over A_0} \left({r\over r_{\rm vir}}\right)^{-\beta} \right]=0\,.
\end{equation}
For convenience, we change the variables to
$x=r/r_{\rm vir}$, $y={\mu m_{\rm p}A_0 \over kT_{\rm vir}}\rho(r)^{\gamma-1}$, 
where the virial temperature of the halo 
$T_{\rm vir}\equiv \mu m_{\rm p} GM_{\rm vir}/2k$.
The equation to be solved is then
\begin{equation}\label{equ:ode}
{{\rm d} y \over {\rm d} x}+{\gamma-1 \over \gamma} 
\left[ \beta {y \over x} + 2  {m(x) \over x^{\beta+2}} \right] = 0\,,
\end{equation}
where $m(x)$ is the halo mass profile normalized by the virial 
mass. For a NFW profile given by Eq.\,(\ref{equ:NFWprofile}),
the normalized mass profile is 
\begin{equation}
m_{\rm NFW}(x) = {\ln(1+x c) - 
{ x c \over 1+ x c} \over \ln(1+c) - { c \over 1+ c}}\,.
\end{equation}

We numerically solve the 1-dimensional differential equation for 
the gas density profile using the Runge-Kutta Dormand-Prince method 
\citep{Dormand1980}. The boundary condition is chosen 
so that the total gas mass enclosed by the virial radius equals
the total hot gas of the halo at the time in question.  
Once the density profile is obtained, the corresponding 
temperature profile can be calculated based on the assumed 
entropy profile as:  
\begin{equation}
T(r) = A_0 \left({r\over r_{\rm vir}}\right)^{\beta} 
{\mu m_{\rm p}\rho(r)^{\gamma-1} \over k}\,.
\end{equation}

In some special cases, the density profile can be solved analytically.
For example, the density profile of an isentropic gas in a NFW halo is 
\begin{equation}
\rho={{\rho}_{\rm vir}} \left\{1+{4\over5}\mathcal{E} 
{c \over \ln(1+c) - { c \over 1+ c}} 
\left[ {\ln (1+cx) \over cx} - {\ln (1+c) \over c}\right]   
\right\}^{3/2}\,,
\end{equation}
where $\mathcal{E}$ is defined as the ratio of the halo virial temperature 
and the gas temperature at the virial radius, $\mathcal{E}\equiv {T_{\rm vir} /T(r=r_{\rm vir})}$.
This solution is similar to that of \citet{Maller2004} 
but generalized for an arbitrary gas entropy value. Similarly, if a dark 
matter halo has a singular isothermal density profile, $\rho_{\rm DM}\propto r^{-2}$, 
the hydrostatic equilibrium gas density profile can be 
written as $\rho\propto r^{-{3\over2}\beta}$ for $\beta \neq0$. 
For the special case where $S\propto r^{4/3}$, the gas density 
profile is then parallel to the dark matter profile as $\rho\propto r^{-2}$ and the temperature profile 
is constant. If $\beta=0$, the gas density profile is 
\begin{equation}
\rho(r)=\rho_{\rm vir} \left[ 1- {4\over5} \mathcal{E} 
\ln {r\over r_{\rm vir}} \right]^{3/2}\,,
\end{equation}
which is equivalent to the solution obtained by \citet{Mo1996}. 

Figure \ref{fig:hot_equ} shows the density and temperature profiles 
for the hot gas in hydrostatic equilibrium in a NFW halo 
of $M_{\rm vir}=10^{12}\msun$ predicted by assuming  different 
entropy profiles, $\beta=4/3, 1, 1/2$ and 0, as indicated in 
the figure. For all the models, the normalization of the 
entropy profile is chosen so that $\mathcal{E}=1$, 
i.e. the temperature at the virial radius equals the halo virial 
temperature. The figure shows that hot halo gas with 
a shallower slope for the entropy profile has more extended gas distribution. 

In the preventative model that we consider in this paper, when the baryonic matter 
is preheated to an entropy level equal to or higher than the virial entropy of the halo, 
the gas accretes into the halo adiabatically. 
Therefore, the hot halo gas will conserve its entropy and will have a flat entropy profile, 
corresponding to an isentropic gas distribution ($\beta=0$).
In contrast, if the baryonic matter is cold and has an entropy much lower 
than the virial entropy of the halo, the collapse of the gas is expected to induce 
a virial shock, which produces a steep entropy profile.  
We will explore the impact of the 
different entropy profiles as a consequence of different assumptions for 
the entropy of the circum-halo matter on the growth of central disks. 

\subsection{Radiative cooling of hot halo gas}
\label{sec:model_cooling}

In many semi-analytic models of galaxy formation, 
cooling time is compared with a pre-defined 
timescale (e.g. the dynamical time of the system)
to determine whether the gas can cool:  
no gas can cool if the cooling time is longer than 
the time-scale chosen. In reality, however, there
must be fluctuations in the density and temperature 
of the hot gas so that part of the hot gas can still 
cool, developing a multi-phase medium, even if
the overall cooling time is long. 
Moreover, when the halo potential has been established in 
the fast-accretion phase, the inner hot 
gaseous halo can cool continuously without depending on 
the newly accreted gas on the halo outskirts. 
To catch the realistic cooling process, we consider a modified model in which 
gas is allowed to cool slowly even if the cooling 
timescale is long. 

In practice, we first choose the values of $A_0$  
and $\beta$ to set up the gas distribution, with 
the gas density profile normalized so that the total 
hot gas mass matches $M_{\rm hot}$ predicted by the model.
We follow the gas density profile with 200 shells equally 
spaced in $\log r$ from $r=0.03\kpc$ to $r=300\kpc$, 
with the gas density and temperature in each shell 
given by the solution of the hydrostatic equilibrium 
equation. We then compute the cooling timescale for each shell, using 
\begin{equation}
\tau_{\rm cool}(r) = {3\over 2} {\mu m_{\rm p} k T \over \rho \Lambda(T, Z)}\,,
\end{equation}
where $\Lambda(T, Z)$ is the cooling function adopted from \citet{Sutherland1993}.
If the cooling timescale is shorter than the 
free-fall time of the shell, $\tau_{\rm ff}=r_i/V_{\rm c}$, where $r_i$ is the radius of the shell 
and $V_{\rm c}$ is the circular velocity of the halo, 
the gas mass in the shell is assumed to cool and be 
accreted onto the central galaxy in a free-fall time.
If, on the other hand, the cooling timescale is 
longer than the free-fall time, the gas in the shell
is assumed to cool and be accreted over the cooling timescale.
In both cases, the accretion rate of the cooling gas
onto the central disk from a mass shell, $i$, is given by 
\begin{equation}
\dot{m}_{i, \rm cool} = { m_{i, \rm hot} 
\over \rm max(\tau_{i, {\rm cool}}, \tau_{i, \rm ff})}\,,
\end{equation}
where $m_{i, \rm hot}$ is the hot gas mass of the mass shell. 
The total gas accretion rate is just a summation of the cooling rates from all radii, 
\begin{equation}
\dot{M}_{\rm cool} 
= \sum_i \dot{m}_{i, \rm cool}\,,
\end{equation}
where the summation is over all the shells that are enclosed by the virial radius of the halo. 

\begin{figure*}
\begin{center}
\includegraphics[width=0.8\textwidth]{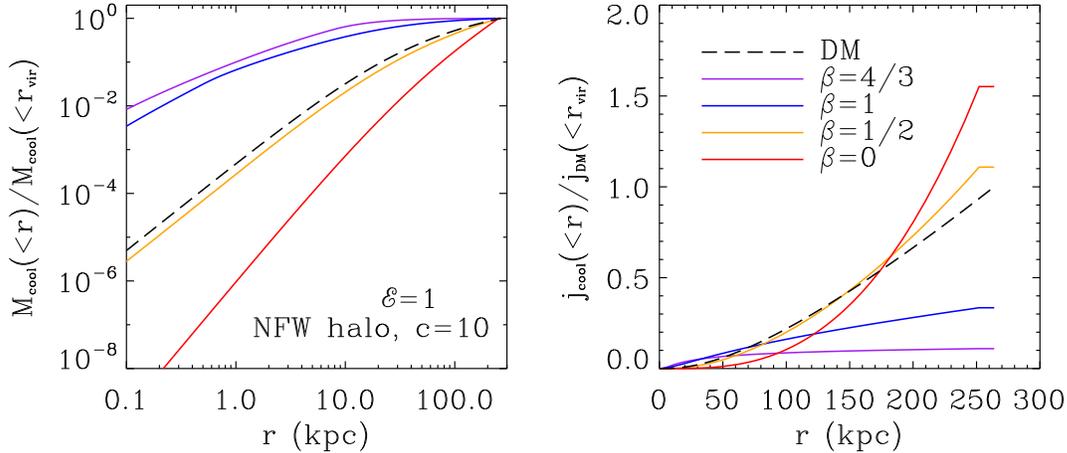}
\caption{Left: the cooling gas mass profile predicted by models with 
different entropy radial profiles. When $\beta$ increases, the 
entropy profile has a steeper entropy profile. $\beta=0$ corresponds 
to an isentropic configuration, which is assumed in the preheating model we explore 
in the present paper. The black dashed line shows the 
dark matter mass profile of the NFW halo with $c=10$. 
Right: the specific angular momentum profile of the cooled 
baryonic matter. The black dashed line shows the specific 
angular momentum profile of dark matter as a function of radius. 
Flatter entropy profiles have more extended density 
distribution and result in more cooling from large radii 
and higher specific angular momentum. 
}
\label{fig:cooling}
\end{center}
\end{figure*}

The left panel of Figure \ref{fig:cooling} shows the cooling gas 
mass as a function of the radius where the gas is cooling from
for different hot gas profiles in a NFW halo.
For comparison, the black dashed line shows the halo mass 
profile, $M(<r)$. As one can see, a steeper entropy profile 
predicts a more concentrated cooling gas mass distribution;  
the isentropic case with $\beta=0$ predicts that the 
cooled gas has a much more extended distribution 
than the dark matter distribution.
The figure illustrates that the thermal state of the circum-halo 
medium has a clear physical consequence on where the halo gas 
can cool and eventually contribute mass and angular momentum to the disk.  

To estimate the total angular momentum of the cooling baryonic matter 
at a given time, we assume that the specific angular momentum of the halo material 
has a radial profile given by  
\begin{equation}
j(r)\propto r^{\alpha}
\end{equation}
with $\alpha=1.1$ \citep[see][]{Bullock2001}.  
The baryons cooling from a given radius are assumed 
to have the same specific angular momentum as the 
dark matter at the same radius. Thus,  the specific angular 
momentum of the cooling baryonic matter relative to that of 
the entire halo is 
\begin{equation}\label{equ:jcool}
{j_{\rm cool} \over j_{\rm DM} }= 
{ \int j(r) {\rm d} m_{\rm cool}(r) \over \int {\rm d} m_{\rm cool}(r)} 
 {\int {\rm d} m_{\rm DM}(r) \over \int j(r) {\rm d} m_{\rm DM}(r) }\,.
\end{equation}

Since the cooling gas from the hot halo gas with a shallower 
entropy profile comes preferentially from larger radii
where specific angular momentum is higher,  
the cooling gas is expected 
to have a larger specific angular momentum than the halo.
We compute the accumulative angular 
momentum of the cooling baryonic matter and normalized it 
by the same quantity of the dark matter. 
The right panel of Figure \ref{fig:cooling} shows the 
normalized specific angular momentum profile as a function of radius 
for halo gas with different entropy profiles.  
For a steep entropy profile, cooling is dominated by inner halo 
where angular momentum is relatively low, and so the resulting 
${j_{\rm cool}/j_{\rm DM}}$ ratio is much smaller than one.  
It has been well appreciated that disk galaxies need to retain most of 
their primordial specific angular momentum 
to match the observed $j_*-M_*$ relation \citep{Dutton2012, Romanowsky2012}.
Our calculation demonstrates that the halo gas with a steep entropy profile unavoidably produces 
disks that are too concentrated to match observations \citep[e.g.][]{Fall2002}.
In contrast, when the entropy profile is flat, 
the baryonic matter can cool from the outer part of a halo where 
the specific angular momentum is larger, and the 
contribution of low angular momentum gas is reduced. 
For the isentropic case ($\beta=0$), which is expected in the preheating model, we find 
${j_{\rm cool}/j_{\rm DM}}\sim 1.5$, which can result 
in a disk with a characteristic radius about 5 times 
as large as that formed in a steep entropy profile with $\beta=1$.  

\subsection{Formation of central disk galaxies}
\label{sec:model_cold}

We assume that, in every timestep, the newly accreted cold gas 
has an exponential radial profile with an angular 
momentum the same as that of the accreted gas, and is added 
to the existing cold gas disk. The exponential scale radius 
is determined using the \citet[][hereafter MMW]{Mo1998} model,
\begin{equation}
r_{\rm d}={\lambda \over \sqrt{2}} f_j f_c^{-1/2} r_{\rm vir}\,, 
\end{equation}
where $\lambda$ is the spin parameter of the gas, 
$f_j\equiv(J_{\rm cool} /m_{\rm cool})/(J_{\rm DM}/M_{\rm vir})={j_{\rm cool}/j_{\rm DM}}$
is the ratio of the specific angular momentum of 
the newly accreted gas and the halo as we compute using Eq.\,(\ref{equ:jcool}), 
$r_{\rm vir}$ is the virial radius of the halo at the time in question,
and $f_c$ is a term depending on the halo density profile.
We adopt Eq. (23) in \citet{Mo1998} to compute $f_c$ for a NFW halo 
density profile. 
In our modeling, the effect of contraction 
of the dark matter halo due to disk formation  
\citep{Blumenthal1986, Gnedin2004, Choi2006} is ignored. 
We assume $\lambda=0.035$, which is the median value 
of dark matter halos in cosmological simulations 
\citep{Bullock2001, Maccio2007}. 
The model implies that the angular momentum is perfectly conserved as 
the gas cools and accretes onto a central disk. 
In addition, we ignore the scatter in the halo spin parameter in this paper. 
When the scatter is taken into account, the scatter for the predicted galaxy size 
will increase. 

When the local surface density of the cold gas on the disk is higher than a certain 
threshold, we assume star formation starts to proceed. 
We adopt a molecular star formation model proposed by \citet{Krumholz2009}. 
The implementation of this model can be found in Appendix \ref{sec:model_sf}. 
The model predicts a surface density of star formation of a disk galaxy 
as a function of radius, $\Sigma_{\rm SFR}(R)$. 
The formed stars return a fraction of mass back into the interstellar medium (ISM) 
as they evolve over time. 
In our model, we implement a time dependent mass return model, which 
is described in Appendix \ref{sec:model_mlost}. 
The model allows us to trace the mass return from star formation in the past over 
the entire evolution of a model galaxy. We treat the mass return as a local process, 
so that the surface density of the mass return rate as a function of 
radius, $\Sigma_{\rm re}(R)$, depends on the star formation history of the annulus at the radius $R$. 

\begin{figure*}
\begin{center}
\includegraphics[width=0.45\textwidth]{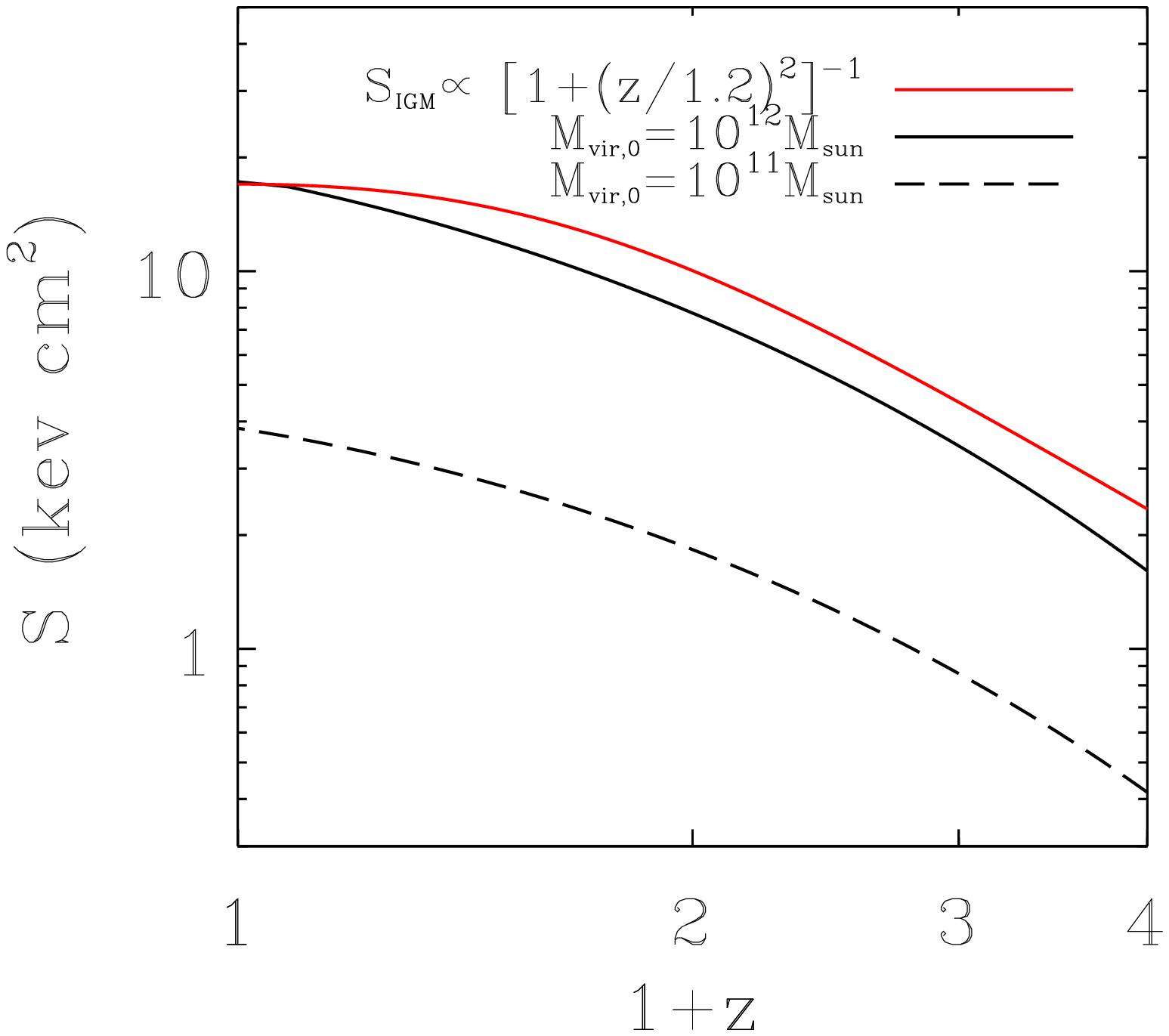}
\includegraphics[width=0.45\textwidth]{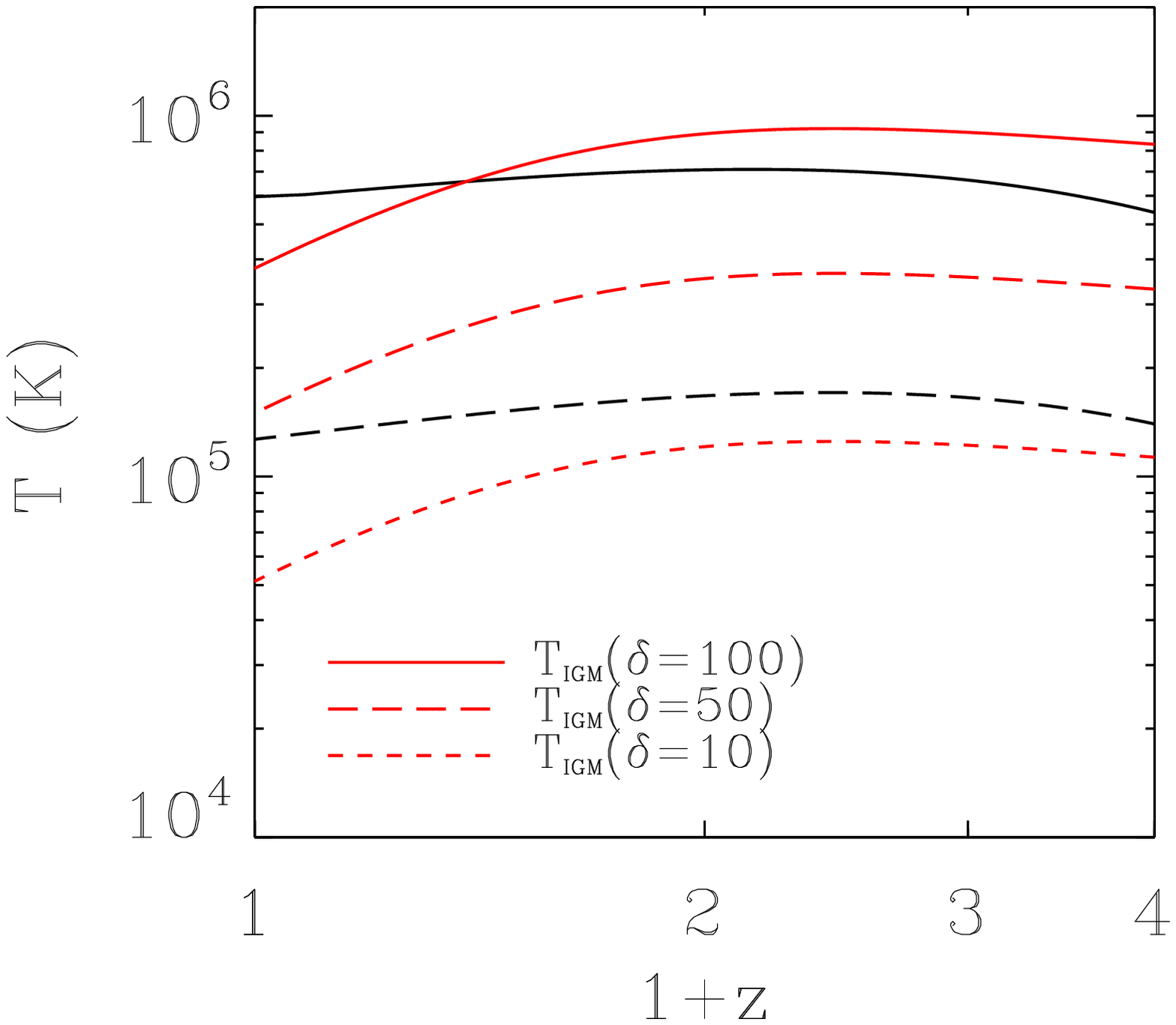}
\caption{The red line in the left panel shows the history of the
assumed preheating entropy as a function of redshift. 
The right panel shows the corresponding temperature of 
the preheated medium with different overdensities. The solid red line 
shows the corresponding temperature if the gas has an over-density $\delta=100$. 
The long and short dashed lines show the corresponding temperature 
if the gas has densities equal to 50 and 10 of the mean density 
of the universe, respectively.
The virial entropy and temperature of dark matter halos with final
($z=0$) masses $10^{12}\msun$ and $10^{11}\msun$ are 
shown by the black solid and dashed lines, respectively, in the panels for comparison. 
The preheating entropy we assume is similar to the virial entropy of present-day $10^{12}\msun$ halos, 
and about 5 times higher than that of present-day $10^{11}\msun$ halos. 
}
\label{fig:preh_hist}
\end{center}
\end{figure*}

The cold gas in the central disk can be also affected by star formation feedback. 
Here we model the effect of such feedback
by taking into account possible ejection of cold gas from 
the galactic disk through outflows. The outflow rate is 
assumed to be proportional to the star formation rate 
with an efficiency $\alpha_{\rm LD}$, known as the mass loading factor. 
Thus, the surface density of the outflow rate of each annulus is written as  
\begin{equation}\label{equ:of}
\dot{\Sigma}_{\rm of}(R)=\alpha_{\rm LD} \Sigma_{\rm SFR}(R)\,.
\end{equation}
Since the timescale of feedback (due to massive stars) is   
shorter than the timescale of mass loss due to stellar evolution,
we treat the feedback as an instantaneous process 
following star formation.  Including the outflow, the evolution 
of cold disk gas is then given by  
\begin{equation}
\dot{\Sigma}_{\rm gas}(R)=\dot{\Sigma}_{\rm cool}(R) - \Sigma_{\rm SFR}(R) - 
\dot{\Sigma}_{\rm of}(R) + \dot{\Sigma}_{\rm re}(R)\,.
\end{equation}
For the ejective feedback model, we maximize the effect of the ejection 
by assuming the ejected mass leaves the halo and is never reincorporated back into the halo.
This is an extreme assumption, but we will show that even with such an extreme 
assumption, the ejective feedback model still tends to over predicts the cold baryon mass
in low-mass halos.

In summary, for a given hot halo, whose 
structure and evolution are modeled in Sections  
\ref{sec:model_hot} and \ref{sec:model_prof}, the value 
of $\dot{\Sigma}_{\rm cool}$ at a given time is obtained from  
the cooling model described in Section \ref{sec:model_cooling}
in combination with the disk model described in Section
\ref{sec:model_cold}. 
Once $\dot{\Sigma}_{\rm cool}$ is predicted and a model for 
$\alpha_{\rm LD}$ is adopted, Equations (\ref{equ:sf}) in Appendix \ref{sec:model_sf}, 
(\ref{equ:freturn}) in Appendix \ref{sec:model_mlost} and (\ref{equ:of}) 
can be combined to solve for the surface density profiles of 
different disk mass components: stars, total cold gas, 
molecular gas, and atomic gas. 
All these together provide a complete
prescription to follow the formation and evolution of a galaxy 
disk in an evolving dark matter halo.

\section{Model Predictions}
\label{sec:results}

Using the model described in the previous section, we make 
predictions for the properties of galaxy disks and their redshift 
evolution. In this paper, we consider two distinctively different 
models to demonstrate the impact of preheating on the 
formation and evolution of disks in halos with a mass 
similar to or lower than that of the Milky Way:
\begin{itemize}
\item {\bf Model-EJ}: an ejective feedback model. 
  No preheating is assumed, i.e. the circum-halo gas has
  an initial entropy that is much lower than the level produced by
  virial shocks ($S\ll S_{\rm vir}$). Nearly the cosmic baryon fraction 
  of baryonic matter is accreted into halos and is shock heated. 
\item {\bf Model-PR}: a preventative feedback 
model assuming preheating, i.e. the circum-halo gas is preheated to have an
  initial entropy that is higher than 
  or comparable to the halo viral entropy ($S\gtrsim S_{\rm vir})$. 
  Ejection is completely switched off at $z<2.5$.   
\end{itemize}

In Model-EJ, because the gas is cold when it is accreted into a dark matter halo,
a virial shock will heat the gas up to the virial
temperature when it is incorporated into the virial radius of the halo. 
For simplicity, we assume that the post shock gas has an 
entropy profile with $\beta=1.1$ in Eq.(\ref{equ:en_prof}), 
consistent with adiabatic hydrodynamical 
simulations and theoretical models for virial shock heating \citep[e.g.][]{Tozzi2001, Voit2005, Lu2007}. 
In this model, we follow the commonly adopted energy-driven 
wind model, which assumes that the mass loading factor of star 
formation feedback is proportional to $V_{\rm c}^{-2}$, 
where $V_{\rm c}$ is the halo circular velocity at any given time. 
Specifically, we assume $\alpha_{\rm LD}=
2\left({200{\rm km\,s}^{-1} \over V_{\rm c}}\right)^2$. 
We note that this choice of outflow mass-loading factor is typical 
for many existing SAMs \citep[e.g.][]{Benson2003, DeLucia2004, Somerville2008}
and, as we will show later, the parameters we adopt produce a model that 
closely matches the baryon mass-halo mass ratio for Milky Way sized galaxies.

In Model-PR, we assume that the circum-halo medium is generally
preheated before it collapses into halos, and the entropy level increases with 
time. As a simple model we take 
\begin{equation}
S =S_0\left({M_{\rm vir, 0} \over 10^{12}\msun}\right)^{\mu}{1 \over 1+(z/z_{\rm c})^{\nu}}\,,
\end{equation} 
where $S_0$ is an amplitude, $z_{\rm c}$ a characteristic 
redshift, and $\mu$ and $\nu$ control the halo mass 
and redshift dependence, respectively. 
The form is chosen  to capture the entropy of the 
IGM generated by various processes discussed in 
the Introduction. 
The entropy is expected to build up over cosmic time, 
and to increase with halo 
mass because at a given time, halos with higher masses
are biased toward higher density regions where star 
formation is more active. We have tried varying 
the parameters, and found that setting 
$S_0=17\,{\rm Kev\,cm^2}$, $z_c=1.2$, $\mu=0.2$ 
and $\nu=2$ matches the observational data remarkably well. 
Figure\,\ref{fig:preh_hist} shows the entropy and 
temperature histories of this preheating model.
If the preheating entropy is equal to or higher than 
the virial entropy of the halos, no strong accretion shock is 
expected as the gas accretes into dark matter halos. 
The hot halo gas is then expected to have a 
flat entropy profile, corresponding to an 
isentropic gas distribution ($\beta=0$). If, on the other hand, the virial 
entropy is higher than the preheating entropy, we neglect
the effect of preheating and the model is equivalent to Model-EJ.

\begin{figure}
\begin{center}
\includegraphics[width=0.45\textwidth]{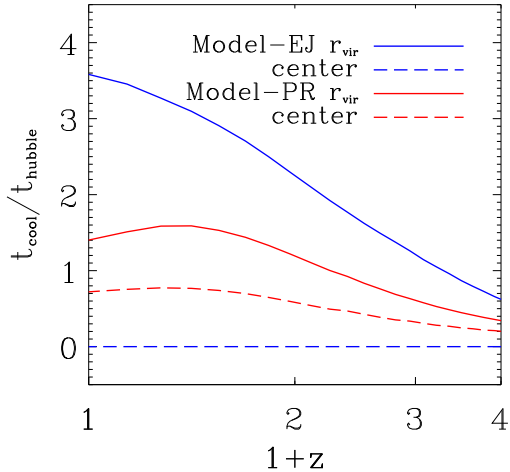}
\caption{The cooling timescale of a halo with a final mass of 
$10^{12}\msun$ at the virial radius (solid line) and at the halo
center ($r=0.03$kpc, dashed line) predicted by Model-EJ (blue) and Model-PR (red) 
as a function of redshift. The cooling time is normalized by 
the Hubble time at the corresponding redshift $z$. 
}
\label{fig:tcool_hist}
\end{center}
\end{figure}

To understand the impact of preheating on the cooling 
timescale of the halo gas, we take a smooth halo mass accretion
history for a halo with final mass $M_{\rm vir,0}=10^{12}\msun$ 
at $z=0$ and compute the cooling timescale of the halo gas at the 
virial radius and the halo center ($r=0.03$kpc) using Model-EJ 
and Model-PR. In the calculation, we switch off cooling and 
star formation and just to show the cooling timescale given by
the original gas distribution predicted by the two models 
under different assumptions of the entropy.  
The results are shown in Figure \ref{fig:tcool_hist}. 
In Model-EJ, the cooling timescale is not only a strong function 
of radius, owing to the steep gas density profile, but also depends  
strongly on redshift. The cooling timescale is always much shorter 
than the Hubble time at the halo center. At the virial radius, the 
cooling timescale is shorter than the Hubble time at high redshift 
when the mass of the progenitor is low, but the cooling time becomes several times 
longer than the Hubble time at low redshift as the virial 
temperature increases and gas density decreases. 
The situation is very different in Model-PR, where the 
cooling timescale in units of Hubble time depends only 
weakly on radius and redshift. The cooling timescale of the 
halo gas stays roughly in the range from 0.5 to 1.5 times
the Hubble time over a large cosmic time since $z=2$. 
Thus, the effect of preheating is to create a situation 
where the cooling time of the halo gas in low-mass halos
is comparable to the Hubble time.       

In the preheating model, no outflow is assumed  
at low redshifts when the star formation rate is reduced 
by preheating. At high redshifts before preheating, 
a constant mass loading factor with no halo mass dependence 
is assumed. 
Specifically, we write the loading factor as
 \begin{equation}\label{equ:feedback_preh}
\alpha_{\rm LD}(z)={\alpha_{\rm LD, inf} \over 2}\left[1+{\rm
    erf}\left({z-z_{\rm c} \over \Delta z_{\rm c}}\right)\right],
\end{equation}
and take $\alpha_{\rm LD, inf}=5$, $z_{\rm c}=2.5$ and 
$\Delta z_{\rm c}=1$. This early outflow only affects star formation at 
high redshift ($z\gtrsim2.5$), which is not the focus of this paper. 

\begin{figure*}
\begin{center}
\includegraphics[width=0.9\textwidth]{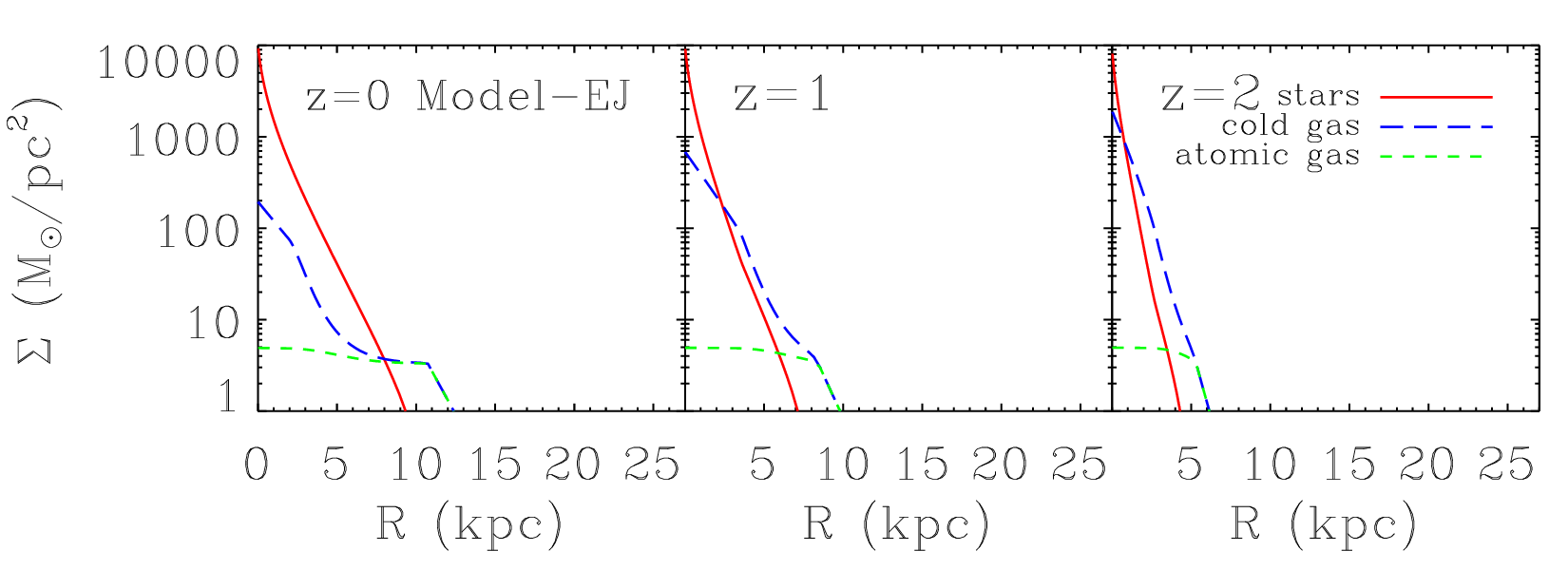}
\includegraphics[width=0.9\textwidth]{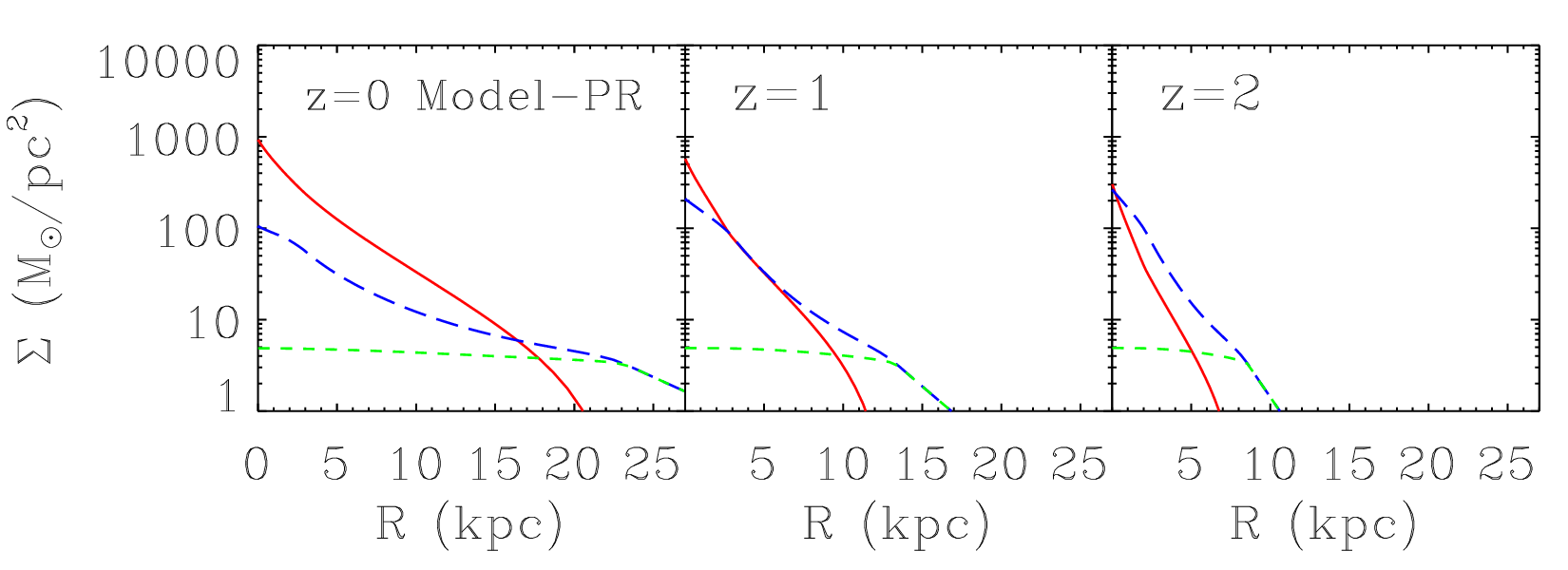}
\caption{The surface density profiles of the cold baryonic matter of 
the central disk at $z=0$, 1 and 2 predicted by Model-EJ (upper panels)
and Model-PR (lower panels).  In each panel, the red solid line 
shows the stellar mass surface density of the central disk 
as a function radius; the blue long-dashed line is the total 
cold (atomic plus molecular) gas mass surface density profile;  
the green short-dashed line is for atomic gas only. 
The preheating model produces significantly more extended disks.
}
\label{fig:disc_prof}
\end{center}
\end{figure*}

\begin{figure*}
\begin{center}
\includegraphics[width=0.9\textwidth]{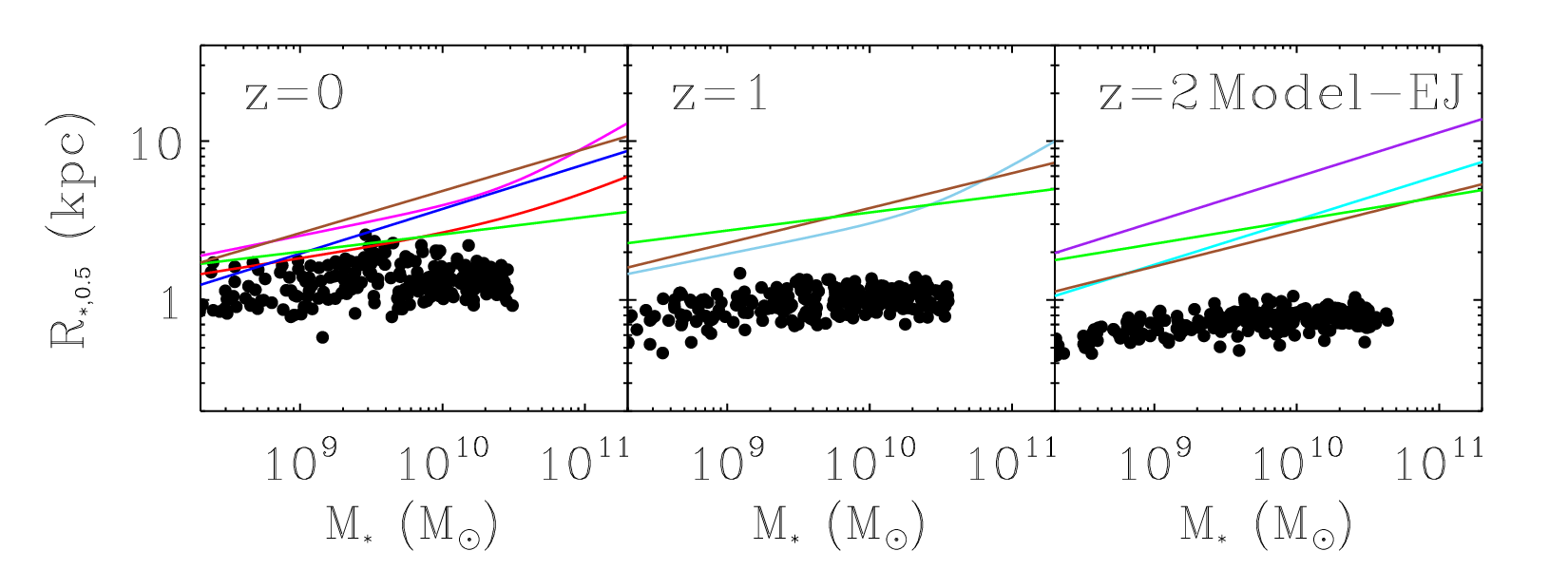}
\includegraphics[width=0.9\textwidth]{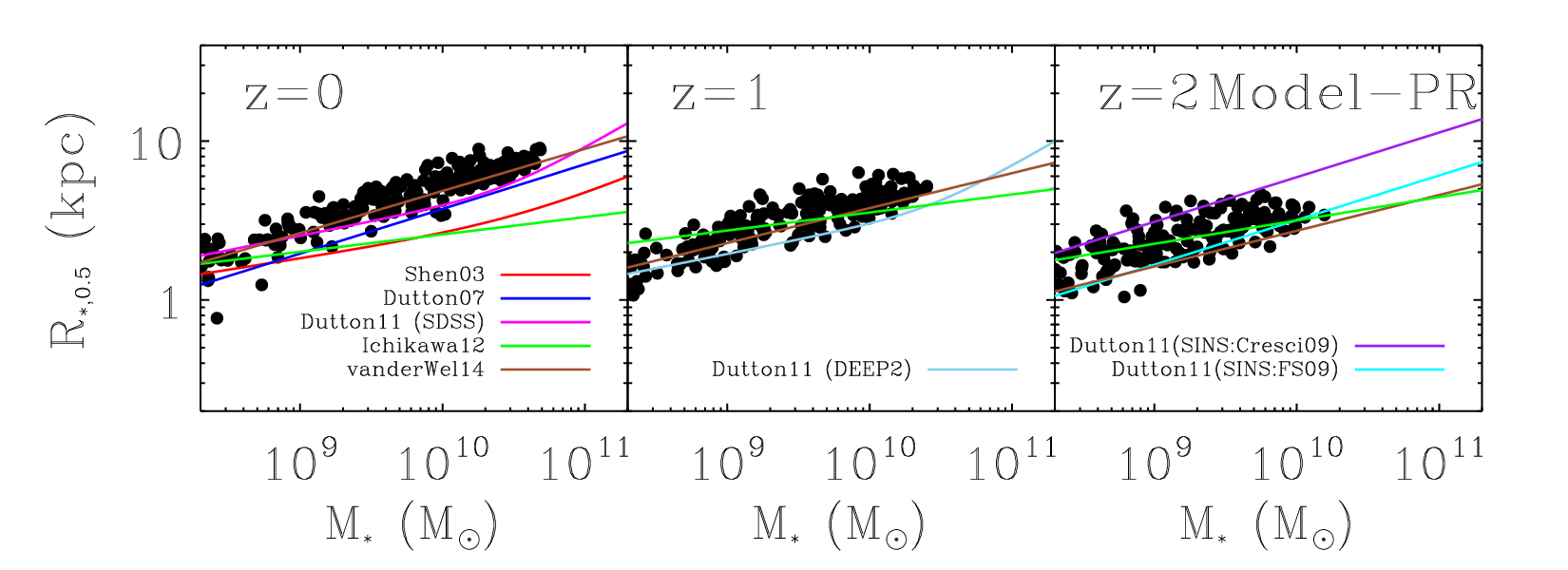}
\caption{Half mass radius of the stellar disk as a function of stellar
mass at $z=0$, 1 and 2.  The black dots are model predictions using
a set of 200 randomly selected MAHs from a cosmological simulation at each redshift. 
All halos have final masses ranging from $10^{10}$ to $10^{12}\msun$ at each redshift.  
The upper row shows the predictions of Model-EJ, and the lower row shows Model-PR.
The lines are compilations of observational data. 
The red line is the result of \citet{Shen2003}, and the blue line is
the result of \citet{Dutton2007} for local galaxies. 
The green lines and brown lines are the results of \citet{Ichikawa2012} 
and \citet{vanderWel2014}, respectively, for the 
corresponding redshift. The light blue line in the $z=1$ panel 
is the result of \citet{Dutton2011} for DEEP2 data. 
The purple and cyan lines in the $z=2$ panel are the results of 
\citet{Dutton2011} based on the results of \citet{Cresci2009} 
and \citet{ForsterSchreiber2009}, respectively, both  
derived from the SINS data. 
}
\label{fig:rd_sample}
\end{center}
\end{figure*}

\subsection{The growth of disks}

Using the two models described above, we predict how a disk builds up.  
Figure \ref{fig:disc_prof} shows the predicted surface density 
profiles of the stellar mass, cold gas mass and atomic gas at 
$z=0 $,  $1$ and 2 for a halo with a final mass 
$M_{\rm vir,0}=10^{12}\msun$. Results are shown 
for both Model-EJ (upper panels) and Model-PR (lower panels).  
In both cases, the stellar mass surface densities follow roughly 
an exponential profile with a density enhancement at the 
center ($r\lesssim3$kpc). The atomic gas has a flat distribution
and dominates  the total cold (atomic plus molecular) gas in the 
outer disk, while the total gas distribution is concentrated and 
dominated by the molecular gas in the inner region of the disk.  
Furthermore, the atomic gas disk is more extended than the 
stellar disk at all redshifts because the outer gaseous disk has
surface densities too low to form molecular gas and stars.  
At any given redshift, Model-PR predicts a more extended disk
than  Model-EJ, because the hot halo gas can cool from a more 
extended volume in Model-PR, and, hence, the disk has larger 
angular momentum. 

\begin{figure}
\begin{center}
\includegraphics[width=0.45\textwidth]{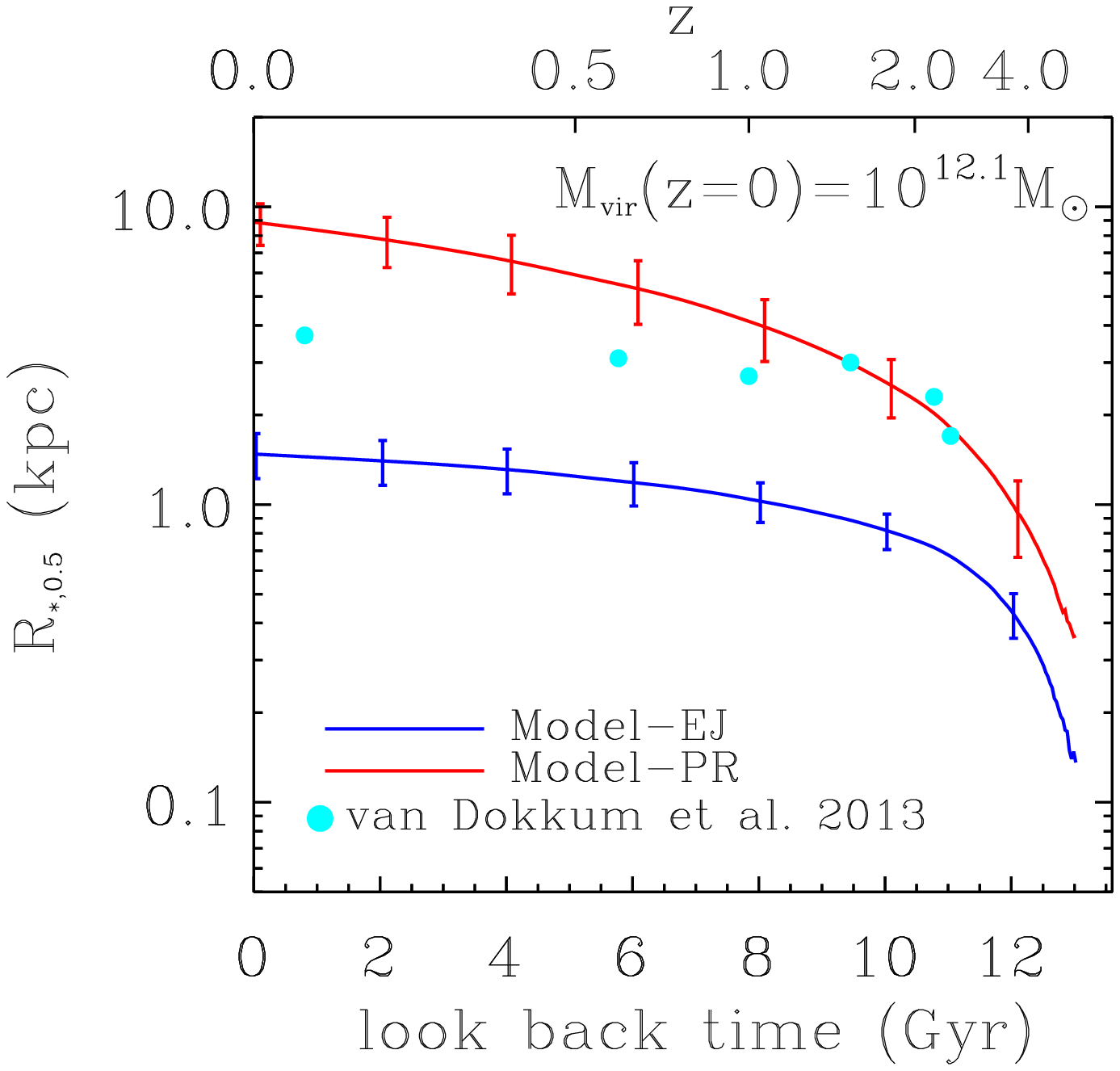}
\caption{The evolution of the half-stellar mass radius of the central 
galaxy of a halo with final mass $10^{12.1}\msun$ at $z=0$. 
The blue line is the prediction of Model-EJ,  and the red line 
is that of Model-PR. 
The filled circles denote the observational result of 
\citet{vanDokkum2013} for the evolution the progenitors of the present-day 
Milky Way size galaxies since $z\sim 2.5$. 
}
\label{fig:hist_rdisc}
\end{center}
\end{figure}

To quantify the structural evolution of disk galaxies,  
we randomly select 200 halos with masses ranging from $10^{10}$ 
to $10^{12}\msun$ from the Bolshoi simulation volume at each of the three redshifts, $z=0$, 1 and 2,  
down-weighting low-mass halos with a selection probability 
$p\propto M_{\rm vir}^{1.3}$ to void having too many low-mass halos that are similar 
in the predicted quantities. We then apply Model-EJ and Model-PR 
to the MAHs of these halos to predict the half stellar mass 
radius, defined as the radius within which half of the disk 
stellar mass is contained,  as a function of stellar mass.
The model predictions are shown in  Figure \ref{fig:rd_sample} 
in comparison with existing observational data.  
For the observational data, we only show the mean relation between 
the half mass radius and stellar mass, as the scatter in the relation is 
not a focus of the model. 
Those observational data include \citet{Shen2003} and \citet{Dutton2007} for local galaxies, 
and \citet{Dutton2011}, \citet{Ichikawa2012}, and \citet{vanderWel2014} for galaxies in a range of redshift 
from $z=0$ to 2.
The upper panels show the predictions of Model-EJ.
We see clearly that this model significantly under-predicts 
the half mass radius, especially at high $z$ and for 
high-mass galaxies. In contrast, the predictions of 
Model-PR, shown in the lower panels, are in excellent agreement with 
the observational data.  

Figure \ref{fig:hist_rdisc} shows the average half mass radii of the
stellar disk as functions of redshift for halos with a present-day
virial mass in a narrow range of $12\leq \log \left({M_{\rm vir,
      0}/\msun}\right) <12.2$ predicted by the two models; 300
randomly selected halo mass accretion histories are used to make this
prediction. The error bars are $1-\sigma$ scatter among the model
galaxies.  The recent observational result of \citet{vanDokkum2013}
for the evolution of the progenitors of the present-day Milky Way size
galaxies since $z\sim 2.5$ is shown in Figure \ref{fig:hist_rdisc} as
filled circles.  We note that the sizes of Milky Way like galaxies in
the \citet{vanDokkum2013} result are roughly two times smaller than
those of disk galaxies in \citet{Dutton2007}, \citet{Dutton2011} and
\citet{vanderWel2014}.  The difference arises mainly because the van
Dokkum et al. sample include all galaxies with the same stellar mass
as the Milky Way, including both star-forming disk galaxies and
quiescent spheroidal galaxies, but other results shown in Figure
\ref{fig:rd_sample} are only for disk galaxies.  In addition,
\citet{vanDokkum2013} measured circularized radii of stacked images,
whereas other results either measure major axis radii or corrected for
inclination.  Knowing these issues, we find that the prediction of
Model-PR is in general agreement with the observational data with an
over-prediction of the size by about a factor of two to three compared
to the observational estimate for Milky Way size galaxies at late
times ($z<1$).  This over-prediction is not surprising because we only
attempt to model disk galaxies, and ignore angular momentum loss,
which is expected in spheroidal galaxies formed mainly through mergers
\citep[e.g.][]{Shen2003}.  This over-prediction leaves room for
angular momentum loss possibly happening in reality.  In contrast,
Model-EJ predicts disk sizes that are too small to match the
observation for all redshifts. If the angular momentum is lost during
the assembly of the disk, the size of the disk is expected to be even
smaller.  In terms of evolution, Model-PR predicts that the half mass
radius increases by a factor of two since $z\sim2$, but the
\citet{vanDokkum2013} result suggests a rather weak evolution.  The
galaxy samples used to represent progenitors of present-day Milky Way
galaxies in \citet{vanDokkum2013} are selected based on a constant
number density.  According to the study of \citet{Behroozi2013c}, the
constant number density selection could overestimate the mass of the
progenitors at higher redshifts.  Given the current uncertainties in
selecting progenitors in observations, the predicted evolution trend
for disk sizes remains to be tested by more accurate observational
estimates.

\subsection{The growth of disk stellar mass}

\begin{figure*}
\begin{center}
\includegraphics[width=0.45\textwidth]{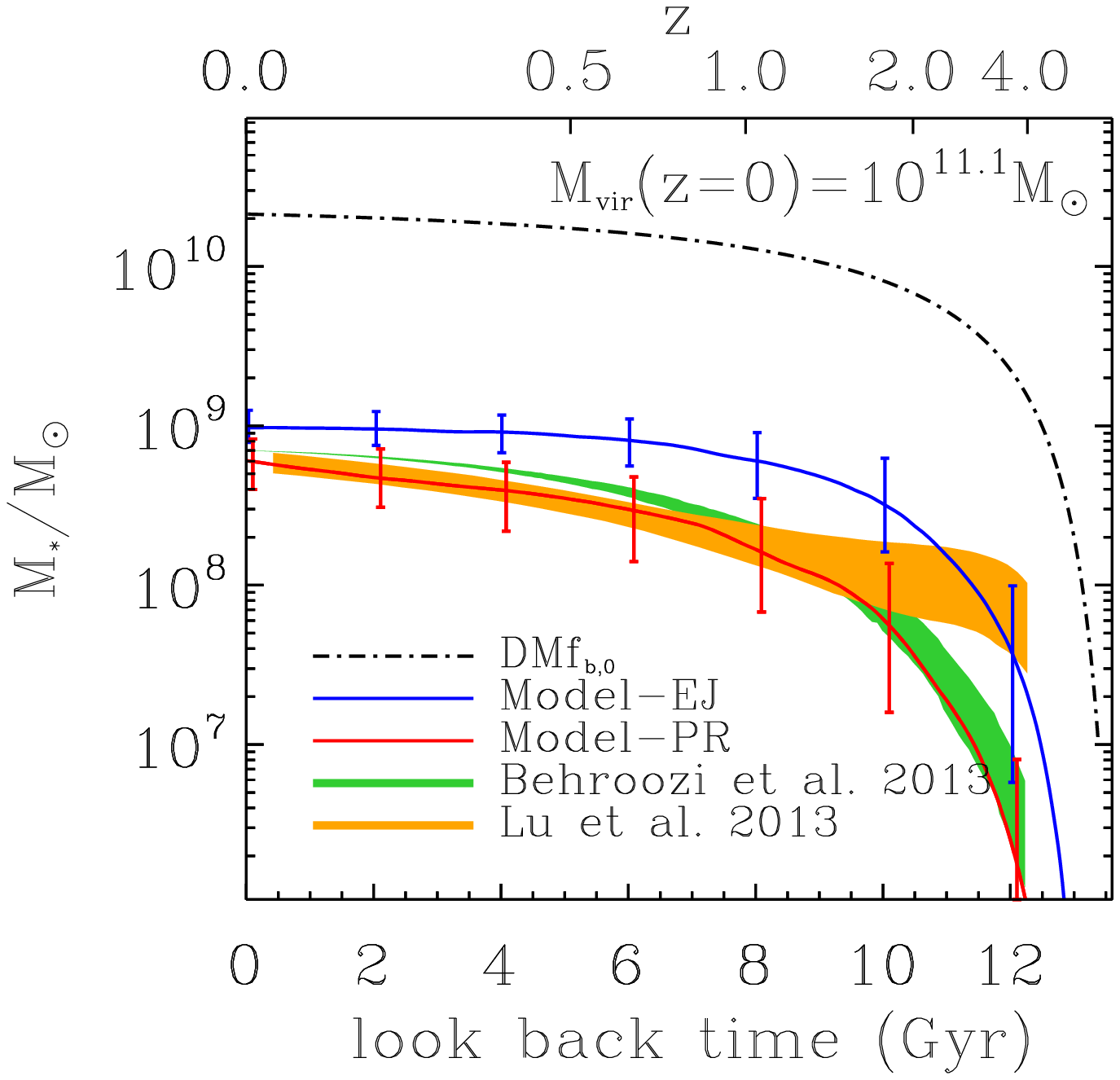}
\includegraphics[width=0.45\textwidth]{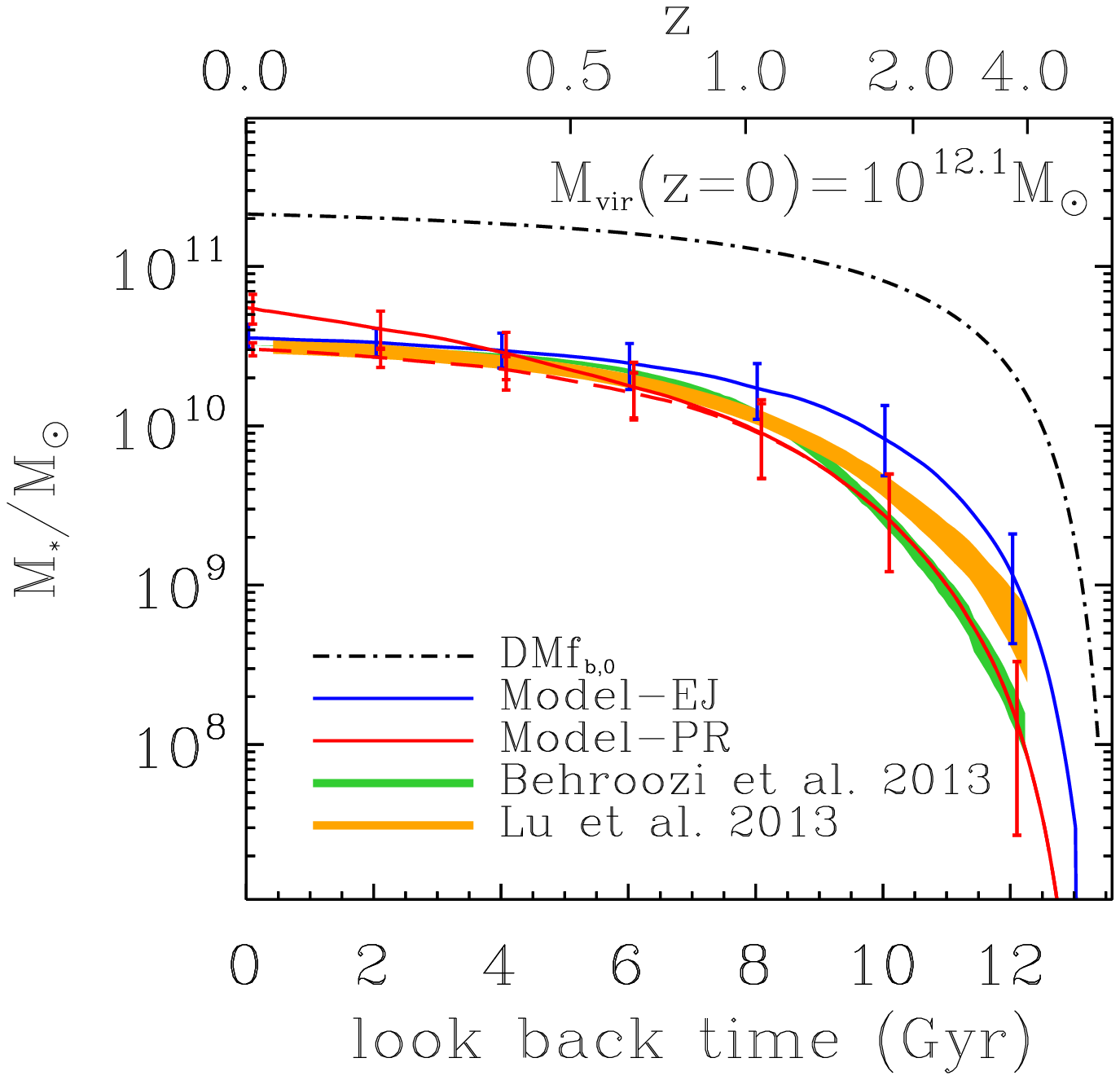}
\caption{The evolution of stellar mass for halos with final mass 
$\sim10^{11.1}\msun$ (left) and $\sim10^{12.1}\msun$ (right) at $z=0$. 
The blue line denotes the prediction of Model-EJ averaged over 200 
simulation halo MAHs, the solid red line denotes the same prediction 
with preheating, Model-PR. The dashed red line denotes a modified 
model based on the preheating model (see text).
The error bars show the standard deviation of the mean 
expected from a sample of 100 galaxies. For comparison the black 
line shows the smoothed halo MAH using Eq.\,(\ref{equ:mah}) 
for the corresponding halo mass multiplied by the universal 
baryon fraction $f_{\rm b,0}$ as a function of time (or $z$). The color bands show the 
central galaxy stellar mass histories of obtained from 
the empirical models of \citet{Behroozi2012} and \citet{Lu2013}. 
}
\label{fig:hist_mstar}
\end{center}
\end{figure*}

In this subsection, we examine how the stellar mass of disks 
grows in halos with final masses of about $10^{11.1}\msun$ and $10^{12.1}\msun$ 
in both Model-EJ and Model-PR. We again use 300 randomly selected
simulation MAHs for each of the two final halo mass bins,
$11\leq \log \left({M_{\rm vir, 0}/\msun}\right) <11.2$ 
and  $12\leq \log \left({M_{\rm vir, 0}/\msun}\right) <12.2$, 
and apply the two models to these MAHs to make predictions. 
Figure \ref{fig:hist_mstar} shows the stellar mass of central 
galaxies as a function of time (redshift), with the left 
panel for the $\sim 10^{11.1}\msun$ 
halos and the right panel for the $\sim 10^{12.1}\msun$ halos. 
The solid lines are the median stellar masses of the MAHs, 
and the error bars show 
the 50\% of the distribution around the median. 
The predictions of Model-EJ and Model-PR are shown as 
the blue and red lines, respectively. The predictions show that
stars form earlier in Model-EJ than in Model-PR, and 
the difference is larger for the lower mass halos.
For $M_{\rm vir, 0}=10^{11.1}\msun$ halos, Model-EJ predicts a stellar 
mass at $z=2$ about 5 times higher than Model-PR. 
Even for $10^{12.1}\msun$ halos, the prediction 
of Model-EJ at $z=2$ is about 2 times as high as 
that of Model-PR at the same redshift.  We compare our 
model predictions with results obtained by 
\citet{Behroozi2012} and  \citet{Lu2013} using 
observationally constrained empirical models. 
The high value of $M_*$ at $z\geq 2$ for 
$M_{\rm vir, 0}= 10^{11.1}\msun$ halos obtained by 
\citet{Lu2013} is due to a boost of star formation 
in low-mass halos at high $z$ as they inferred  
to match the faint-end luminosity function of present-day 
clusters of galaxies. 
We do not attempt to capture this behavior in our model
because the total mass of stars formed in this 
mode is only a small fraction ($\sim 1/10$) of the total final 
stellar mass, and it happens at high redshift ($z\gtrsim2$). 
As one can see,  Model-EJ over-predicts the stellar 
mass  for $M_{\rm vir, 0}= 10^{11.1}\msun$ 
halos over the entire redshift range,  while 
Model-PR matches the empirical results remarkably well 
given the simplicity of the model. 
For $M_{\rm vir, 0}= 10^{12.1}\msun$ halos,  
Model-EJ matches the empirical results well
at $z<1$ but still over-predicts $M_*$ at higher $z$.
In contrast, Model-PR matches the empirical 
results reasonably well at $z>1$ but over-predicts 
$M_*$ at very low redshift ($z<0.5$). 

\begin{figure*}
\begin{center}
\includegraphics[width=0.45\textwidth]{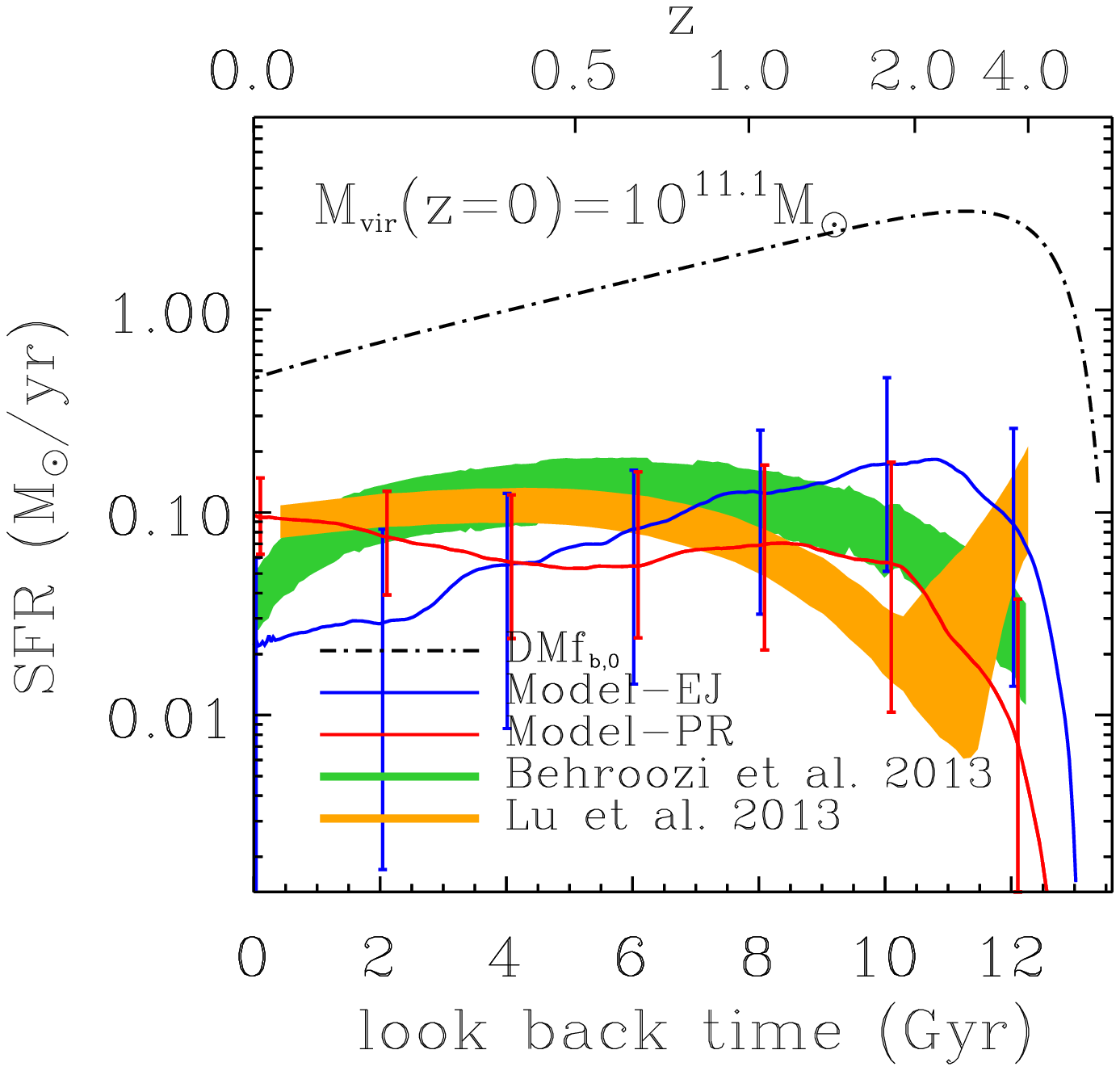}
\includegraphics[width=0.45\textwidth]{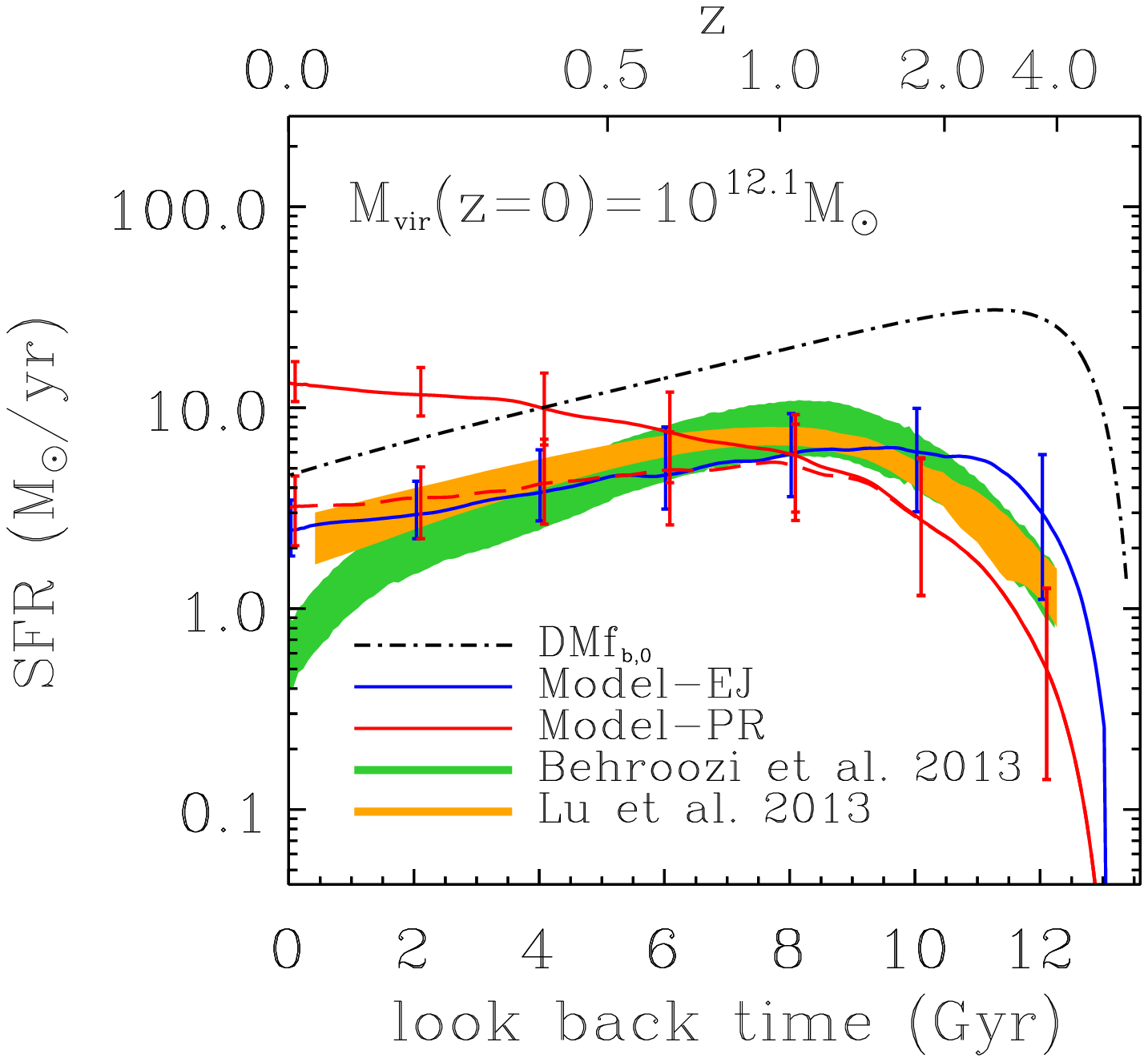}
\caption{The SFR histories of halos with final 
mass $10^{11.1}\msun$ (left) and $10^{12.1}\msun$ (right) at $z=0$. 
The blue line is the prediction of Model-EJ, the solid red line is 
prediction of Model-PR, and the dashed red line is the prediction
of a modified model based on the preheating model, 
which will be discussed \S\ref{sec_latequenching}. 
The error bars are the expected standard deviation of the mean with 100 galaxy samples. 
The black line denotes the halo mass accretion rate characterized by Eq. \ref{equ:mah} 
multiplied by the universal baryon fraction $f_{\rm b}$ as a function of $z$. 
The color bands show the central galaxy stellar mass histories 
obtained  by \citet{Behroozi2012} and \citet{Lu2013}. 
}
\label{fig:hist_sfr}
\end{center}
\end{figure*}

These discrepancies between the model predictions and 
the empirical results can also be seen in the star formation 
rate histories shown in Figure \ref{fig:hist_sfr}. 
For $M_{\rm vir, 0}=10^{11.1}\msun$ halos, Model-EJ predicts a SFR history 
that is peaked at $z\approx2$, in contrast to the rather flat histories
predicted by Model-PR and obtained from the empirical models. 
For the case of $M_{\rm vir, 0}=10^{12.1}\msun$, 
although Model-EJ reproduces the decreasing trend for SFR at late time as seen in 
the empirical results, it predicts a broad peak for the star formation history at $z\sim 3$, 
which is much earlier than that in the empirical results.  
Overall, the ejective feedback implemented in Model-EJ is more effective 
in suppressing star formation in low-mass halos and at late times. 
This trend makes it difficult for the model to match of the SFR histories derived from the empirical models. 
In contrast, Model-PR generally predicts a rising SFR history for both mass bins. 
It matches the empirical results well for the lower mass bin, and early times ($z>1$) 
for the $10^{12.1}\msun$ halos.
At late times, however, the model produces a slowly rising SFR history. 
If the observed decreasing SFR for the $10^{12.1}\msun$ halos at low-$z$ is real, 
it suggests that some of the galaxies in this halo mass bin must be undergoing 
some quenching process at late times and this process occurs in relatively massive 
galaxies but not low-mass ones. We will come back to a possible implication of 
this behavior in \S\ref{sec_latequenching}. 

\begin{figure*}
\begin{center}
\includegraphics[width=0.45\textwidth]{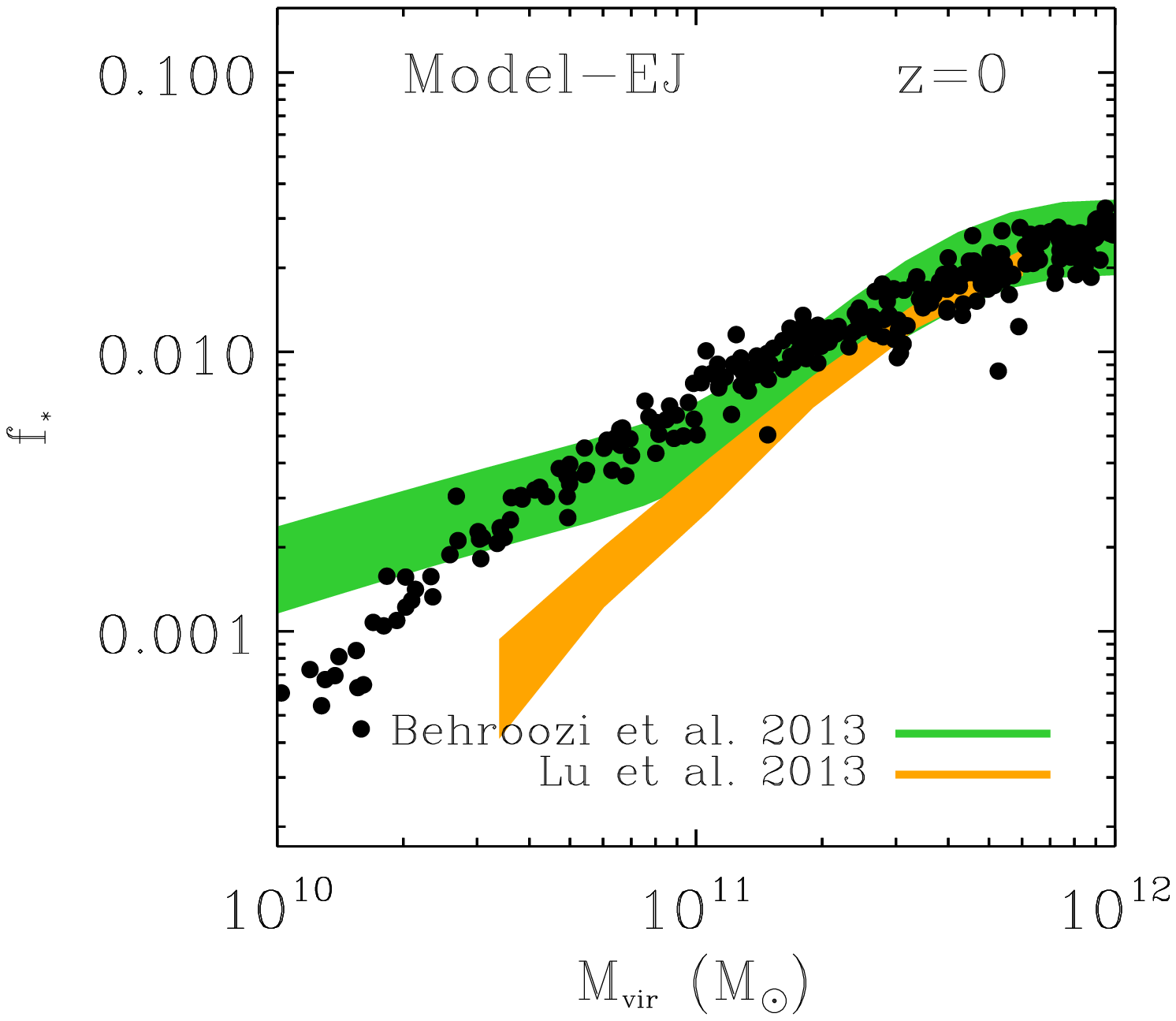}
\includegraphics[width=0.45\textwidth]{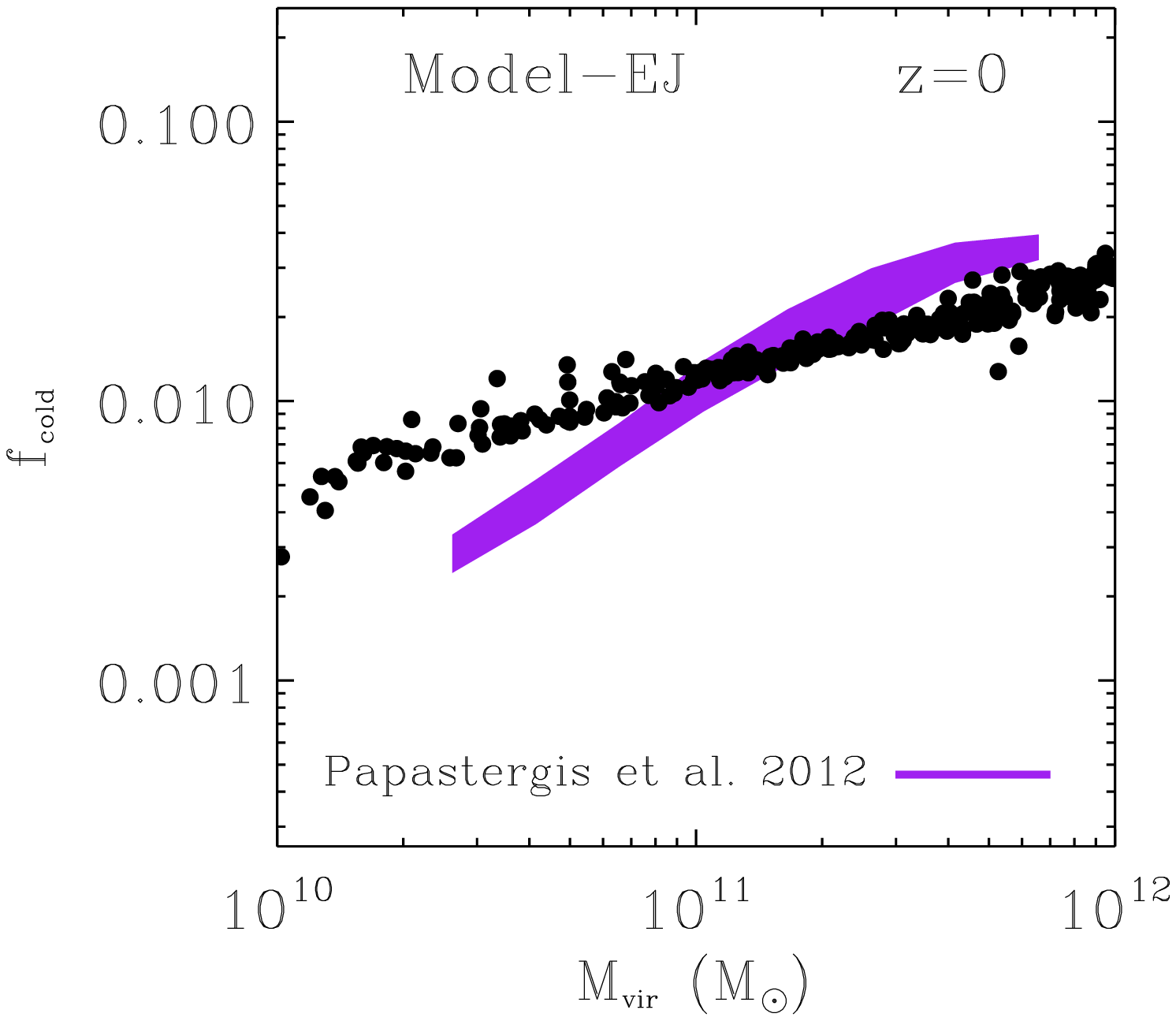}
\includegraphics[width=0.45\textwidth]{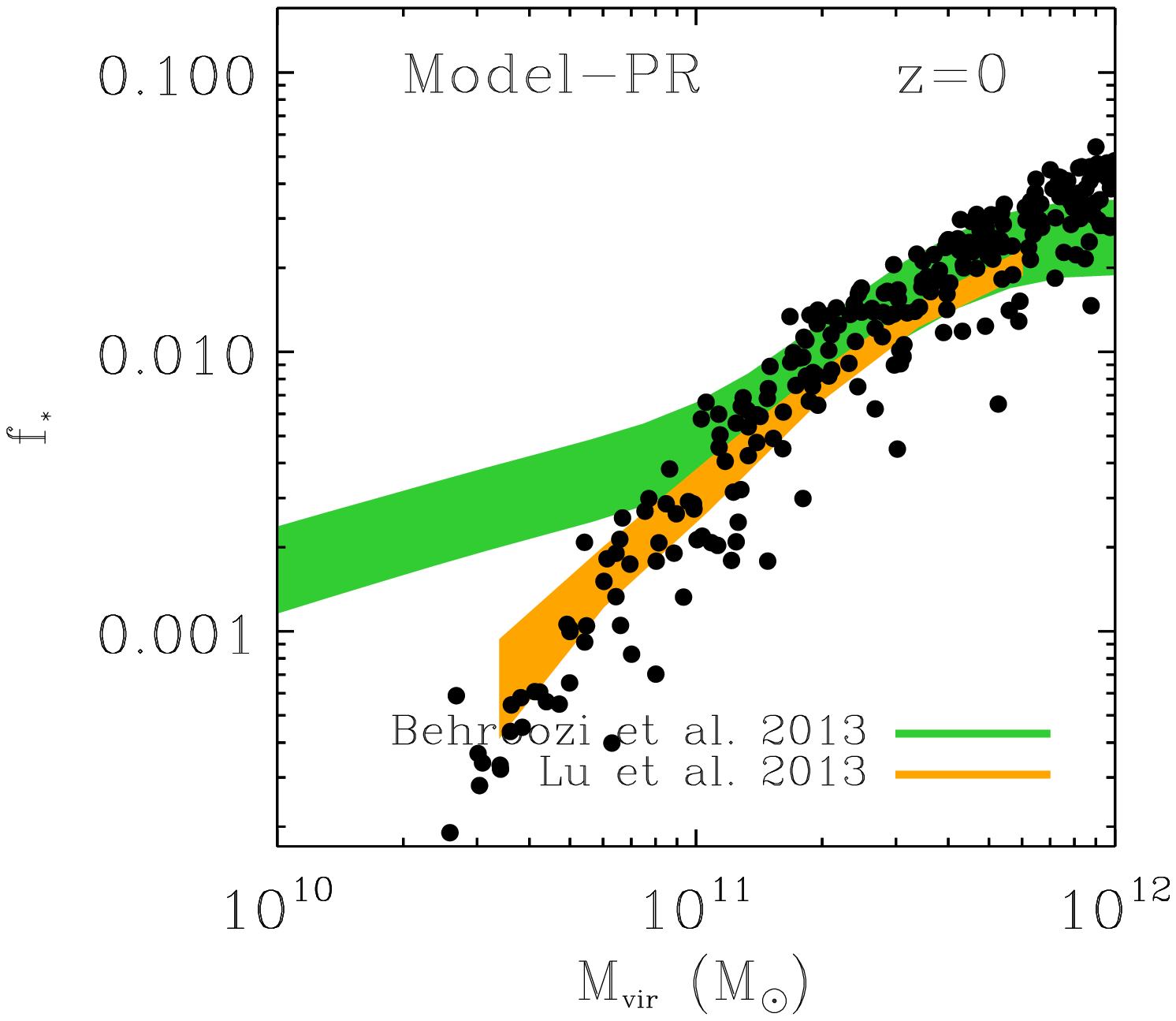}
\includegraphics[width=0.45\textwidth]{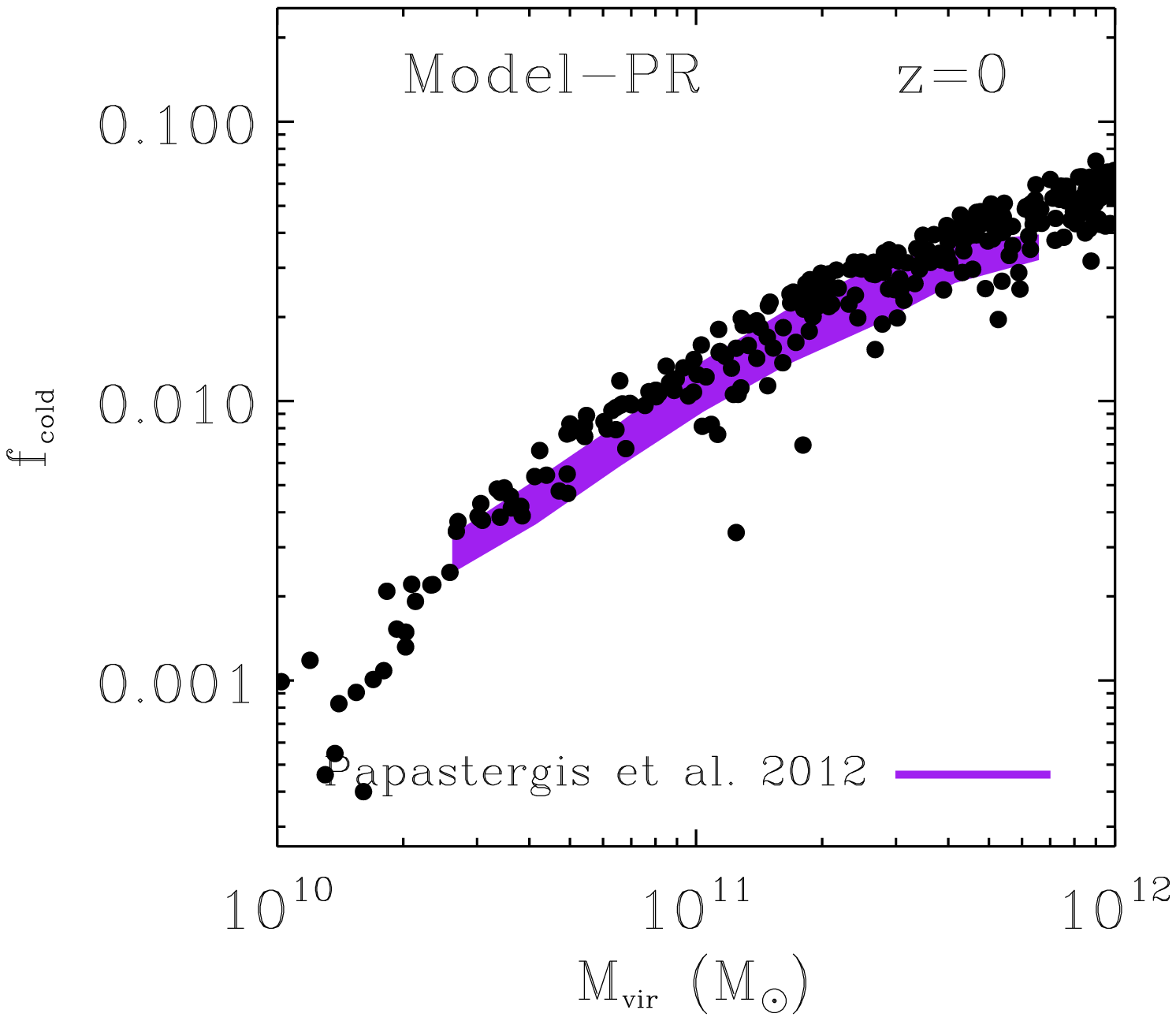}
\caption{The baryon mass  fractions as a function of halo mass for
 central galaxies at $z=0$. The black dots show 
the model predictions for 200 randomly selected realistic halo accretion histories.  
The left column shows the stellar mass fraction, $M_*/M_{\rm vir}$, predicted by the models
and compares the predictions with the results of \citet{Behroozi2012} 
and \citet{Lu2013} derived from observational data (the differences between the two are largely 
due to uncertainties in the faint-end slope of the stellar mass function). 
The right column shows the ratio between stellar mass plus atomic cold gas mass and halo mass. 
The predictions are compared with the results of \citet{Papastergis2012}. 
The upper row panels are for Model-EJ, and the 
lower row panels are for Model-PR. 
}
\label{fig:fm_z0}
\end{center}
\end{figure*}

\subsection{Cold baryon mass fractions in dark matter halos}

We use 200 randomly chosen halo MAHs with final mass ranging from 
$10^{10}\msun$ to $10^{12}\msun$ at $z=0$ to make 
predictions for the final stellar mass and the cold gas mass in both 
the atomic and molecular phases of the central galaxies hosted 
by those halos.  In Figure \ref{fig:fm_z0}, we show the predictions for 
the stellar mass fraction (stellar mass to halo mass ratio) as a function of halo mass 
at $z=0$ and compare the predictions with results of 
\citet{Behroozi2012} and \citet{Lu2013} in the left panels. 
In the right panels, we show the cold baryon mass 
(stellar mass plus atomic cold gas) fraction as a function of halo mass and compare 
the model predictions with the result of \citet{Papastergis2012} for the same 
quantities derived with the abundance matching technique. 

As the feedback parameters in Model-EJ are tuned to match the observed
stellar mass and cold gas mass for $10^{12}\msun$ halos, 
it is not surprising that the predictions of the model at the
high-mass end is in agreement with the data. 
However, the model predicts a much shallower slope of the 
cold baryon mass--halo mass relations. 
If we trust the cold baryon mass fraction result of \citet{Papastergis2012} for halo masses lower than 
$10^{11}\msun$, Model-EJ seems to over-predict the cold baryon mass fractions for low-mass halos, 
even though this model invokes strong feedback outflow. 
For the particular model we adopt, 
the mass-loading factor is as large as 11 for halos with mass $10^{11}\msun$ at $z=0$. 
The result demonstrates that in order to further reduce the cold baryon 
mass fraction in low-mass halos, the model based on the assumption of 
full baryon accretion and strong outflow would 
require an even stronger ejection and an even larger mass-loading 
factor for low-mass galaxies than what we adopt here. 
In contrast, Model-PR nicely reproduces 
the decreasing cold baryon mass fractions for both the stellar 
mass and the cold gas over the entire halo mass range.
This is because the uniform entropy assumed for the cirum-halo gas naturally 
results in a scaling relation that the baryon mass fraction in a halo is proportional 
to the halo mass. 
The predictions of the two models suggest that accurate measurement for 
the stellar mass and cold baryon mass fraction for halos with mass lower than $10^{11}\msun$ 
will directly discriminate between the models. 

\subsection{Evolution of the number density of low-mass galaxies}

\begin{figure}
\begin{center}
\includegraphics[width=0.45\textwidth]{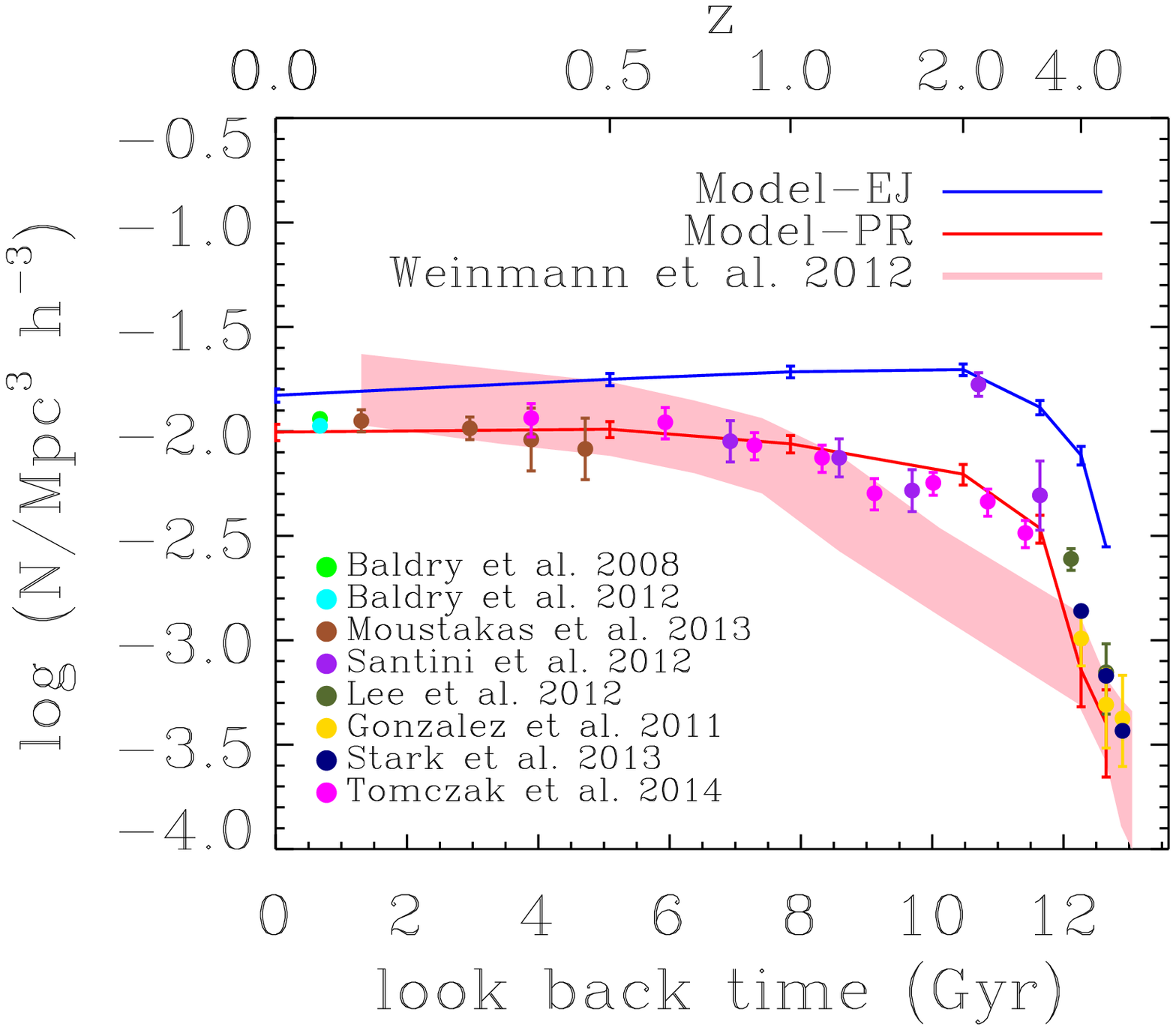}
\caption{The evolution of the number density of galaxies with stellar 
mass in the range  $9.27<\log M_*/\msun<9.77$.
The pink band shows the compilation of data by \citet{Weinmann2012}. 
The color points with error bars are data from recent observational results. 
The sources of the data are listed in the figure.  
}
\label{fig:hist_nden}
\end{center}
\end{figure}

We apply both Model-EJ and Model-PR to mass accretion histories of 
halos with masses in the range of $10^{10}\msun$ and $10^{14}\msun$ 
at $z=0$, $0.5$, $1$, $2$, $3$, $4$, $5$, and $6$, and select
predicted galaxies with stellar masses in the range of 
$9.27<\log M_*/\msun<9.77$ at each redshift. Figure \ref{fig:hist_nden} shows the
predicted number density of galaxies in this stellar mass range 
as a function of redshift in comparison with observational data. 
The pink band shows the data compiled in \citet{Weinmann2012}, 
and the points with error bars are our own compilation
from various recent observations: \citet{Baldry2008} and \citet{Baldry2012} for 
local galaxies; \citet{Moustakas2013} for galaxies out to $z=1$; 
\citet{Santini2012} and \citet{Tomczak2013} for galaxies out to $z\approx3$; 
and \citet{Lee2012}, \citet{Gonzalez2011} and \citet{Stark2013} for
galaxies at higher redshifts. The Santini et al. results give a higher 
number density of galaxies in this stellar mass range than 
the data sets adopted in Weinmann et al. In particular, the data 
of Santini et al. show a very steep low-mass end slope at 
$z\approx2$, leading to an exceptionally high number density 
at this redshift. The number density evolution predicted   
by Model-EJ is very similar to the predictions of the 
SAMs and hydrodynamical simulations shown in \citet{Weinmann2012}
and that of the SAM in \citet{Lu2013a} which is also based 
on the assumption of strong outflow. 
Our result here, therefore, reinforces the general trend expected from 
ejective models assuming full baryon accretion and strong outflow. 
In contrast, Model-PR predicts 
a much lower number density of galaxies in this mass 
range at high redshifts (by half a dex at $z\geq2$). 
The prediction of Model-PR agrees better with data compiled in \citet{Weinmann2012}
than Model-EJ, and it agree with the observational points we compiled 
remarkably well in the entire redshift range, except the jump
at $z\approx2$ in the \citet{Santini2012} data. The difference between the 
model predictions and the discrepancy among the 
observational data clearly demonstrate the importance of 
more accurate determinations of the number density  
of low-mass galaxies in the redshift range of 1 to 4. 
We expect that final analysis of the CANDELS data \citep{Grogin2011, Koekemoer2011} 
will improve these constraints in the near future. 

\subsection{Late quenching of star formation in high-mass galaxies}
\label{sec_latequenching}

Observations have shown that the specific star formation rate, 
which is defined as the star formation rate divided by the stellar mass, 
generally decreases with increasing stellar mass \citep[e.g.][]{Salim2007, Noeske2007}. 
For local galaxies with a stellar mass as high as $10^{10}\msun$, 
about 70\% of them have a specific star formation rate below a certain threshold, 
and hence are classified as quiescent galaxies \citep{Moustakas2013}.
As the analyses of \citet{Lu2013} and \citet{Behroozi2012} represent
the average behavior of star formation histories of the entire 
observed galaxy population, the quiescent population is 
included in their results.  
The models we have considered so far are for star forming 
galaxies, without taking into account any process that may 
quench star formation in high-mass galaxies.
Recent observations have indicated that the quenching of 
star formation in galaxies are closely related to the stellar mass 
surface density in the inner 1kpc square of galaxies 
\citep[][]{Cheung2012, Fang2013a}:
galaxies with a central stellar mass surface density in the central 1kpc square 
area lower than $\sim10^9\msun\,{\rm kpc}^{-2}$ are dominated 
by star forming galaxies, while those with a central stellar 
mass surface density higher than this are quenched.
To mimic such quenching, we include an outflow recipe 
in the preheating model by assuming the mass-loading 
factor depends on the stellar mass as 
$\alpha_{\rm LD}=\left({M_* \over 2.5\times10^{10}\msun}\right)^{2}$, 
so that the feedback only affects high-mass 
galaxies whose central stellar mass density is high.
Although the underlying physics is not well 
understood, this phenomenological model is able to capture
the observational results of \citet{Cheung2012} and \citet{Fang2013a}. 
Indeed, in our model the characteristic disk size scales with  
stellar mass roughly as $r_{\rm d}\sim M_*^{0.3}$.
Assuming all disks have a similar functional form for the stellar surface density 
profile, one finds that the central stellar mass 
surface density goes with the stellar mass as 
$\Sigma_{*,0}\sim M_*^{0.4}$.  Thus the mass-loading factor
assumed above scales with the central stellar 
mass surface density as $\Sigma_{*,0}^5$. This implies 
a sharply increasing feedback strength with an increasing central stellar 
mass surface density, similar to the 
``central surface density'' quenching found in the 
observations in \citet{Cheung2012} and \citet{Fang2013a}.  As shown in Figure 
\ref{fig:hist_sfr} with the dashed lines, 
this model effectively reduces the SFR in high mass galaxies 
since $z\sim 1$, while the SFR history for low-mass 
galaxies is not affected at all. With such a quenching, 
the preheating model reproduces well the decreasing trend of the 
star formation rate at late times for Milky-Way size 
galaxies, but all other predictions are not affected 
significantly (and hence are not shown for the sake of clarity).  
We stress that this stellar mass dependent quenching recipe does 
not fix the problem of Model-EJ, because with strong outflow the model already predicts 
a declining SFR history for Milky-Way sized galaxies, and 
including this quenching recipe makes the decline even more 
rapid and hence more inconsistent with the data. 

\section{Conclusion and Discussion}
\label{sec:conclusion}

In this paper, we have developed a semi-analytic galaxy 
formation model with a self-consistent treatment for the hot halo gas 
configuration and disk formation. 
The model follows realistic halo mass accretion 
histories extracted from a cosmological $N$-body simulation, and 
makes predictions for the structure of baryonic matter in different phases for central disk galaxies 
hosted by halos with mass similar to or lower than the Milky Way galaxy. 
We contrast many predictions made by two models based on different assumptions for 
the thermal state of the circum-halo medium and how the medium is accreted into dark matter halos 
to establish a gaseous halo. The models certainly have uncertainties in various 
components governing the baryonic processes, and we are aware of degeneracy 
between model parameters. Here, we only attempt to demonstrate very basic behaviors 
of the model based on two distinct assumptions. We defer comprehensive analysis 
in the parameter space to a future paper using our established method \citep{Lu2011a}. 

One of the models makes the conventional assumption that the accretion
of gas by halos is from a cold medium.  In this model, baryons
collapse with dark matter and accretion shocks heat the gas to form a
gaseous halo with a steep power-law entropy profile.  This model thus
predicts that cooling is inside out.  In general the disks from such
models contain too much material with low angular momentum, resulting
in galaxy disks that are too compact.
Although, with a fine tuning of 
the feedback mass loading factor, 
the model is able to reproduce the baryon mass fraction of
Milky Way size halos at the present day, it still over-predicts the baryon mass
fraction in low-mass halos, even if the chosen outflow is much
stronger than what is observed.  The model tends to suppress star
formation too much at late times and in low-mass halos, which results
in predicted star formation histories which differ significantly from
current observations. It also predicts too many low-mass galaxies at
high $z$. All these results confirm recent findings of similar
problems in this class of models by other investigators
\citep[e.g.][]{Dutton2012, Weinmann2012, Lu2013a}; the problems do not
seem to be solved by tuning the outflow parameters.

The new model proposed here assumes that the gas to be
accreted by halos has a finite entropy gradually increasing with time
and reaching $\sim 15 {\rm Kev\,cm^2}$ at $z=0$.  The thermal pressure
of the gas with the entropy is high enough to prevent a large fraction
of baryons from collapsing into low-mass halos, and the collapsed gas
has a flat entropy radial profile in the halo, which prevents cooling
from the very central region of the halo.  In this model, gas cooling from
large radii brings higher angular momentum, resulting in a central disk
extended enough to match the observed galaxy size--stellar mass
relation over a large redshift range.  This model also reproduces the
observed relations between stellar mass and halo mass, between cold
baryon (stellar plus cold gas) mass and halo mass, and between disk
size and stellar mass, as well as the observed evolution of the number
density of low-mass galaxies, without invoking excessive outflows at
late times.

In addition to the predictions we have compared with observational
data, the model we propose in this paper has a number of other
observational implications which may be used to test the basic picture. The
preventative model implies that a large fraction of baryons has never
collapsed into low-mass halos.  Furthermore, because all (low-mass)
halos started with relatively low baryon fraction after preheating, no
strong feedback is needed to expel baryonic gas from halos at late
time. Thus, the preheating model predicts that any outflows associated
with low-mass disk galaxies at low-$z$ are weak.  We note, however,
that if feedback from star formation/AGN activities is responsible for
the entropy generation, strong outflows are still expected at high
$z$.  Metals can be tracers, although maybe strongly biased, of inflow 
and outflow. Metallicity measurements of low-mass galaxies may provide 
useful constraints to the model \citep[e.g.][]{Zahid2012}. 
Moreover, the preheated media at the present time implied by our
model may be detected by QSO absorption line systems of highly ionized
elements, such as CIV and OVI, and may indeed be the `missing' baryons
we are looking for \citep[e.g.][]{Yao2008, Gupta2012}.  Finally, our
model suggests that the morphologies of galaxies are closely tied to
the thermal state of hot halo gas and even the circum-halo medium, as
it determines the angular momentum of the baryons that assembly the
disk galaxies.  Such a framework may be used to connect observed disk
properties with underlying dark matter halos, and thereby infer the
spacial distribution of the pre-collapsed baryonic matter
\citep[e.g.][]{Kassin2012}.

The model we explore in this paper is based on a hypothesis of preheated 
circum-halo medium. Although the general picture we capture in this model 
is motived by physical consideration of various plausible early feedback processes, 
the origin of the preheating entropy and detailed physics of how the preventative 
feedback works remain to be solved. 
Because of the unknown physics, the parameters adopted here are largely 
{\it ad hoc}, and the uncertainties of those parameters and their impact on the model predictions 
can not be meaningfully discussed.  
The remaining question is, of course, what is the origin of
preheating? There are a number of suggestions in the literature, 
as described in the Introduction, but none of them has been 
investigated in detail. These different scenarios are expected 
to make different predictions for the level of preheating, as 
well as for its halo mass and redshift dependences. The 
consequences of these differences can be explored in more 
detail using a full semi-analytical model, such as the one 
developed in \citet{Lu2011a}. 
Furthermore, the uncertainties of the baryon processes including 
the preheating process can be constrained by observational data using 
the inference approach developed by \citet{Lu2011a, Lu2012, Lu2013a}. 
We will come back to this in a forthcoming paper.   

\section*{Acknowledgement}
The authors thank Tom Abel, Eric Bell, James Bullock, Edmond Cheung, Aaron Dutton,
Sandra Faber, Zhankui Lu, Ari Maller, Joel Primack, Rachel Somerville, Joop
Schaye, and Frank van den Bosch for useful discussions. 
YL and RHW received partial support from HST-AR-12838, provided by
NASA through a grant from the Space Telescope Science Institute, which is
operated by the Association of Universities for Research in Astronomy,
Inc., under NASA contract NAS5-26555.
HJM acknowledges support from NSF AST-1109354.

\bibliographystyle{apj}
\bibliography{/Users/luyu/references/general}
\appendix

\section{Implementation of the model for reionization}
\label{sec:model_reionization}

We model the effect of photoionization heating by the UV background 
following \citet{Gnedin2000} who showed that the fraction of 
baryons that can collapse into a halo of a given mass can be 
described in terms of a `filtering mass', $M_{\rm F}$. The baryon 
fraction in halos with masses lower than $M_{\rm F}$ is reduced 
relative to the universal fraction according to 
\begin{equation}\label{equ:reion}
f_{\rm b}(z, M_{\rm vir})= 
{f_{\rm b,0} \over [1+0.26 M_{\rm F}(z)/M_{\rm vir}]^3},
\end{equation}
where $M_{\rm vir}$ is the halo virial mass. 
The filtering mass depends on the re-ionization 
history of the Universe and is redshift-dependent. 
\citet{Kravtsov2004} provided fitting formulae for the 
filtering mass according to both the redshift at which the 
first HII regions begin to overlap ($z_{\rm overlap}$) 
and the redshift at which most of the medium is 
re-ionized ($z_{\rm reion}$). We make use of the fitting 
functions (B2) and (B3) in the appendix B of \citet{Kravtsov2004} 
to compute the initial fraction of baryons, $f_{\rm b}$,
as a function of halo mass and redshift. In this paper, we assume 
$z_{\rm overlap}=11$ and $z_{\rm reion}=10$, as suggested 
by WMAP results \citep[e.g.][]{Spergel2007}. 

\section{Implementation of the star formation model}
\label{sec:model_sf}

To predict star formation in a cold gas disk, we adopt the 
model developed by \citet{Krumholz2009}, which 
assumes that the star formation efficiency depends on the 
local total (atomic and molecular hydrogen) gas surface 
density, $\Sigma_{\rm cold}$, and the molecular gas 
fraction, $f_{\rm H_2}$. Specifically, the star formation 
rate surface density is given by
\begin{equation}\label{equ:sf}
\Sigma_{\rm SFR}= \left\{ \begin{array}{ll}
\epsilon f_{\rm H_2} \Sigma_{\rm cold} 
(\Sigma_{\rm cold} /\Sigma_0)^{-0.33},  & \Sigma_{\rm cold} < \Sigma_0;\\
\epsilon f_{\rm H_2} \Sigma_{\rm cold} (\Sigma_{\rm cold} /\Sigma_0)^{0.33},  
& \Sigma_{\rm cold} \geq \Sigma_0,
\end{array}\right.
\end{equation}
where $\epsilon=0.39{\rm Gyr}^{-1}$ and 
$\Sigma_0=85 \msun {\rm pc}^{-2}$.
For a annulus with a surface density of cold gas 
$\Sigma_{\rm cold}$, we compute the molecular fraction, 
$f_{\rm H_2}$, using the model of \citet{Krumholz2009a}:  
\begin{equation}  
f_{\rm H_2} = 
\begin{cases}
1- {3 \over 4} \left({s \over 1+0.25 \xi}\right) & \text{if } \xi<2\,; \\
0 & \text{if } \xi \geq 2\,, 
\end{cases}
\end{equation}
where 
\begin{equation}
\xi = { \ln (1+0.6 \chi + 0.01 \chi^2) \over 0.6 \tau_c}\,,
\end{equation}
in which 
\begin{equation}
\chi=3.1 \left({ 1 + 3.1Z_0^{0.365} \over 4.1}\right)\,,
\end{equation}
and 
\begin{equation}
\tau_c=0.066 \mathcal{C} Z_0 \Sigma_0,
\end{equation}
with $Z_0$ the metallicity of the gas in units of solar metallicity, $Z_\odot$, 
and  $\mathcal{C}$ a clumpness factor that accounts for smoothing 
of the surface density on scales larger than that of a 
single molecular complex \citep[see also][]{McKee2010}.
As suggested in \citet{Krumholz2009}, 
$\mathcal{C} \sim 5$ when $\Sigma_{\rm cold}$ 
is measured on $\sim 1$ kpc scales. In our model 
we take $\mathcal{C}=5$.

\section{Time dependent stellar mass loss model}
\label{sec:model_mlost}

\begin{figure}
\begin{center}
\includegraphics[width=0.45\textwidth]{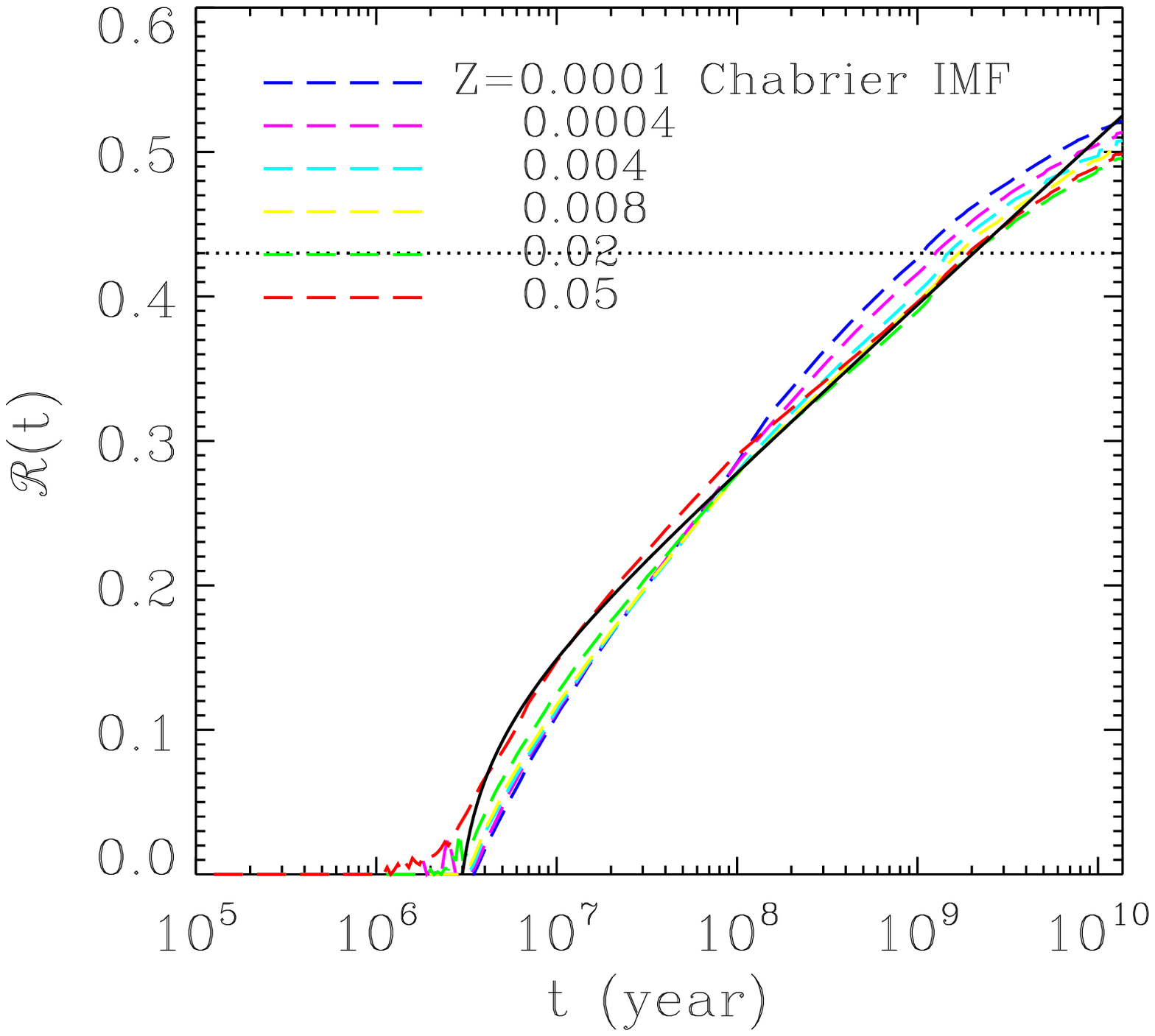}
\caption{The fraction of stellar mass loss due to the evolution of a
  simple stellar population. The dashed lines are predicted by the
  BC03 model for a Chabrier IMF (2003) with different metallicities as
  indicated in the panel.  The black solid line is the model we adopt
  to mimic the stellar mass loss for a Chabrier IMF (see
  Eq.\ref{equ:freturn}).  The dotted line denotes the typical value
  adopted for instantaneous recycling approximation for the IMF.  }
\label{fig:freturn}
\end{center}
\end{figure}

In a time interval $\Delta t$, the stellar mass at a given 
radius in the disk can change due to star formation 
and the mass loss of formed stars, 
\begin{equation}
\dot{\Sigma}_* (R, t) \Delta t
= \Sigma_{\rm SFR}(R, t) \Delta t- \dot{\Sigma}_{\rm re}(R, t) \Delta t\,.
\end{equation}
Here, $\dot{\Sigma}_{\rm re}(R, t) \Delta t$ is the surface density of the stellar 
mass that formed in the past but returned into the ISM in the 
time interval $\Delta t$ at time $t$, and its rate, 
$\dot{\Sigma}_{\rm re}$, is determined by the star formation history and the IMF. 
We write the mass return of the stellar mass formed 
at radius $R$ with a star formation rate $\Sigma_{\rm SFR}(R, t')$ at
an early epoch $t'$ as 
\begin{multline}
\dot{\Sigma}_{\rm re}(R, t, t') \Delta t=  \\
 \left[\mathcal{R}(t-t'+0.5\Delta t)  -\mathcal{R}(t-t'-0.5\Delta t)\right]  \Sigma_{\rm SFR}(R, t')\,,
\end{multline}
where $\mathcal{R}(t)$ is the returned fraction determined by the IMF and stellar evolution.  
We use the following fitting formula proposed by \citet{Jungwiert2001} to describe  
$\mathcal{R}(t)$:
\begin{equation}
\mathcal{R}(t)=\left\{ \begin{array}{ll}
0 & t<t_0; \\
c_0 \ln \left( {t - t_0\over \tau} + 1\right) & t\geq t_0\,.
\end{array}\right.
\end{equation}
The parameters, $c_0=0.05$, $t_0=3\times 10^6\,{\rm yr}$ 
and  $\tau=3.76\times 10^5\, {\rm yr}$ are obtained for a 
\citet{Chabrier2003} IMF. In Figure \ref{fig:freturn} 
we compare this fitting function (the solid curve)
to the return fractions obtained from the BC03 model 
\citep{Bruzual2003} with the same IMF for 
different metallicities (dashed curves). Clearly
the above formula is a good approximation for the 
return fraction over a large range of metallicity.   
Finally we can write 
\begin{multline}\label{equ:freturn}
\dot{\Sigma}_{\rm re}(R, t) \Delta t= \\
 \int_0^t \left[\mathcal{R}(t-t'+0.5\Delta t)-\mathcal{R}(t-t'-0.5\Delta t)\right]
\Sigma_{\rm SFR}(R, t') {\rm d}t'\,.
\end{multline}
This model allows us to trace the mass return over a  
long timescale that is relevant for stellar evolution over 
the entire history of a galaxy.

\end{document}